\makeatletter
\newcommand*{\rom}[1]{\expandafter\@slowromancap\romannumeral #1@}
\makeatother

\documentclass[twocolumn]{aastex631}

\usepackage{amsmath}
\usepackage{gensymb}
\usepackage{hyperref}
\let\tablenum\relax
\usepackage[separate-uncertainty=true]{siunitx}

\begin{document}

\title{Global Coronal Magnetic Field Estimation Using Bayesian Inference}

\author[0000-0001-7816-1857]{Upasna Baweja}
\affiliation{Aryabhatta Research Institute of Observational Sciences, 263001, Nainital, India}
\affiliation{Mahatma Jyotiba Phule Rohilkhand University, Bareilly- 243006, Uttar Pradesh, India}

\author[0000-0002-6954-2276]{Vaibhav Pant}
\affiliation{Aryabhatta Research Institute of Observational Sciences, 263001, Nainital, India}

\author[0000-0002-7008-7661]{I\~nigo Arregui}
\affiliation{Instituto de Astrof\'{\i}sica de Canarias, E-38205 La Laguna, Tenerife, Spain} 
\affiliation{Departamento de Astrof\'{\i}sica, Universidad de La Laguna, E-38206 La Laguna, Tenerife, Spain}

\begin{abstract}

Estimating the magnetic field strength in the solar corona is crucial for understanding different physical processes happening over diverse spatio-temporal scales. However, the high temperatures and low density of the solar corona make this task challenging. The coronal magnetic field is too weak to produce a measurable splitting of the spectral lines using the Zeeman effect, and high temperature causes spectral lines to become weak and broad, making it difficult to detect the small Zeeman splitting. Coronal magneto-seismology, which combines the theoretical and observed properties of magnetohydrodynamic (MHD) waves, can be used to infer the magnetic field strength of oscillating structures in the solar corona, which are otherwise difficult to estimate. In this work, we use the Doppler velocity and density data obtained from the Coronal Multichannel Polarimeter (CoMP) on 2016 October 14 to obtain the global map of the coronal magnetic field using Bayesian inference. Two priors are used for plasma density, viz Gaussian and uniform distributions. Bayesian inference provides us with the probability distribution for the magnetic field strength at each location from 1.05 to 1.35 $R_\odot$. A comparison between the magnetic field obtained using simple inversion and Bayesian inference is also drawn. We find that the values obtained using simple inversion do not always match the maximum posterior estimates obtained using Bayesian inference. We find that the inferred values follow a power-law function for the radial variation of the coronal magnetic field, with the power-law indices for simple and Bayesian inversion being similar.

\end{abstract}

\keywords{The Sun (1693) --- Magnetohydrodynamics (1964) --- Solar coronal waves (1995) --- Bayesian Statistics (1900)}

\section{Introduction} \label{sec:intro}
Different layers of the solar atmosphere are coupled by the magnetic field \citep{jess2016solar}. Thus, the information on the magnetic field is essential to understanding many physical processes of the solar corona, such as coronal heating. The magnetic field estimation of the solar corona is challenging primarily for two reasons. First, the magnetic field in the solar corona is three orders of magnitude lower than in the solar photosphere. This small magnetic field strength produces a minimal Zeeman's splitting of the spectral lines. Second, the million-degree temperature of the solar corona broadens the spectral lines. \cite{1969SoPh....9..131A}, and \cite{1969SoPh....6..442S} proposed the potential free source surface (PFSS) model, an extrapolation method to derive the coronal magnetic field strength. This model was later refined by \cite{1992ApJ...392..310W}. Different extrapolation methods: potential free \citep{Lee_1999}, linear force-free field and non-linear force-free field \citep{2000SoPh..195...89Y,2007MmSAI..78..126R,Su_2009,2011JGRA..116.1101H,2016NatCo...711837X,Huang_2018} are widely used to reconstruct the coronal magnetic field. These models utilize the force-free condition by assuming that the solar corona is current-free or that the currents are not influencing the global magnetic field structure. Such assumptions are not always valid. For instance, the solar magnetic field is expected to be potential, where the dominant force is the magnetic force with no magnetic helicity, and the field must be in the minimum energy configuration. But, these conditions are violated in the active regions, where the magnetic fields are highly sheared or twisted. Also, different extrapolation methods could lead to different results \citep{Lee_1999,2007A&A...468..701R}. \par

Coronal seismology is another method to estimate the magnetic field in the solar corona \citep{1983Natur.305..688R,2005LRSP....2....3N,aschwanden2006coronal,2007SoPh..246....3B,2012RSPTA.370.3193D}.
% is a technique to determine the physical parameters of the solar corona by matching the observations of waves with the MHD wave theory. 
Determination of physical parameters of solar corona such as magnetic field was first suggested by \cite{1970PASJ...22..341U,1970A&A.....9..159R,1984ApJ...279..857R}. However, this idea could not be put into practice until the launch of the Solar and Heliospheric Observatory \citep[SoHO;][]{1995SoPh..162....1D} and the Transition Region and Coronal Explorer  \citep[TRACE;][]{1999SoPh..187..229H}. These spacecraft localized and identified coronal loops, prominences, standing and propagating magnetohydrodynamic (MHD) waves in the solar corona and the transition region \citep{1999Sci...285..862N,1999ApJ...520..880A}. Soon after, the magnetic field strength in the solar corona was estimated, using different ground- and space-based instrumentation by e.g, \cite{1999ApJ...520..880A, 2002SoPh..206...99A,doi:10.1063/1.1324949,2001A&A...372L..53N} using TRACE, \cite{2008A&A...487L..17V}, using Hinode \citep[][]{2007SoPh..243....3K}, \cite{2011ApJ...736..102A, sarkar2016transverse,2016NatPh..12..179J}, using the Solar Dynamic Observatory \citep[SDO;][]{2012SoPh..275....3P} and \cite{doi:10.1126/science.1143304,2017A&A...603A.101L,2020ScChE..63.2357Y}, using the Coronal Multichannel Polarimeter \citep[CoMP;][]{2008SoPh..247..411T}. 

These observations help us infer the average magnitude of the magnetic field in these structures. Extracting the information of the physical parameters, such as magnetic field, from the observations is an inversion problem. To infer parameters from such inversion problems, one has to deal with ill-posed mathematical problems where the number of unknowns may exceed the number of observables. Moreover, these observables cannot be measured with high accuracy. For inversion problems, the use of Bayesian analysis is increasing rapidly in solar physics \citep{Arregui_2011,2012ASSP...33..159A, 2012ASPC..456..121A,2014IAUS..300..393A, 2015RSPTA.37340261A, Scherrer_2017,2017A&A...600A..78P, 2018AdSpR..61..655A, article, 2019A&A...625A..35A,2019A&A...622A..44A,2020ApJ...905...70P,2021ApJ...915L..25A,10.3389/fspas.2022.826947, Pascoe_2022}. The utilization of Bayesian inference in coronal seismology is driven by the need for probabilistic conclusions due to the presence of uncertainties and limited information available. This makes the Bayesian analysis method a more appropriate approach compared to other methods that provide a single numerical estimate. \par
The global map of magnetic field strength was recently inferred using simple inversion by \citet{2020ScChE..63.2357Y}. 
In this study, we inferred the global map of the coronal magnetic field using Bayesian inference on the same data set used by \citet{2020ScChE..63.2357Y}. A comparison is also made between the magnetic field obtained using Bayesian inference and the magnetic field obtained using simple inversion. We also obtained the radial variation of the coronal magnetic field using simple inversion and Bayesian inference. Magnetic field variation along and across the coronal loops have also been obtained and it is found that the coronal magnetic field strength is more inside the coronal loops as compared to the ambient medium. Additionally, the magnetic field strength is found to be more near the loop footpoints than at the loop apex.  \par 
This paper is structured as follows. Section \ref{sec:Instrument} describes the data used. The global map of the coronal magnetic field using simple inversion is obtained in Section \ref{sec:simple}. The magnetic field strength in the solar corona is estimated using Bayesian inference, and the magnetic field's radial distribution is described in Section \ref{sec:Bayesian}. Section \ref{summary} outlines the summary.
\begin{figure*}[t]
    \gridline{\fig{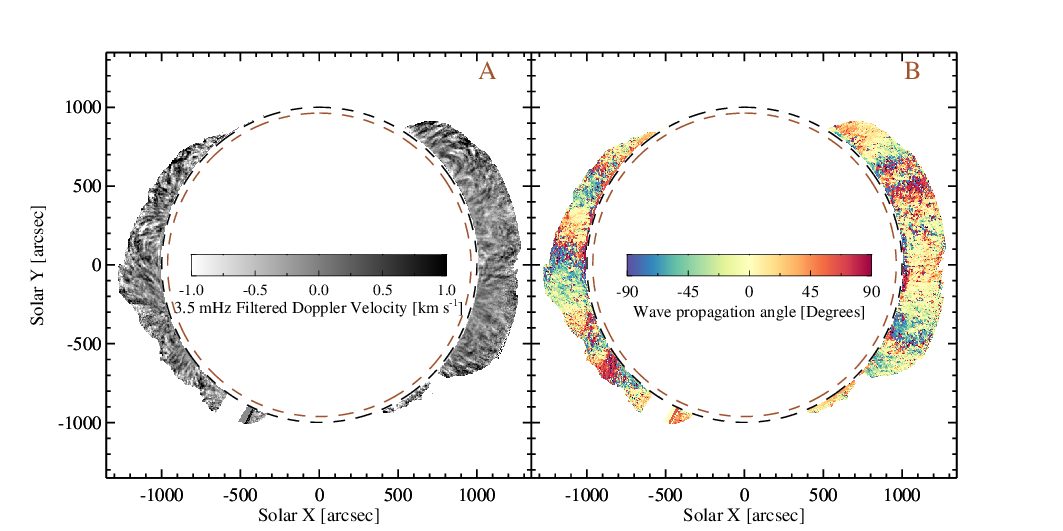}{\textwidth}{}}
    \caption{Images of the solar corona: (A) Map of Doppler velocity of the Fe \rom{13} 1074.7 nm line at 20:39:09 UT on 2016 October 14. A 3.5 mHz Gaussian filter has been applied to the Doppler shift image sequence. An animated version of this panel is available, featuring the 3.5 mHz Gaussian-filtered Doppler velocity from 20:39:08 UT to 21:26:38 UT on the same day. The animation has a real-time duration of 10 seconds. (B) Map of the derived wave propagation direction. In all panels, the brown dashed circle marks the edge of the solar disc (solar limb), and the black dashed circle indicates the inner boundary of the CoMP FOV. The X and Y coordinates represent spatial positions in the east-west and south-north directions, respectively.}
    \label{fig:doppler velocity and magnetic field}
\end{figure*}
\begin{figure*}[b!]
    %\centering
    % \includegraphics[width=\textwidth]{image_1.eps}
    \gridline{\fig{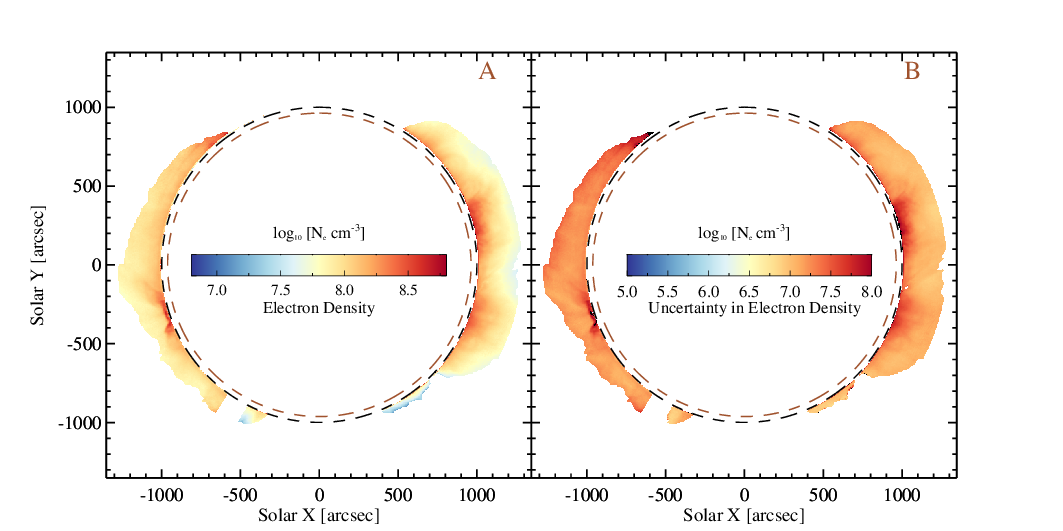}{\textwidth}{}}
    \caption{Density diagnostics (A and B): Maps of the electron density and the associated uncertainty. The circles are as in Figure $\ref{fig:doppler velocity and magnetic field}$. }
    \label{fig:density and its error}
\end{figure*}
\begin{figure*}[b]
    \centering
     \gridline{\fig{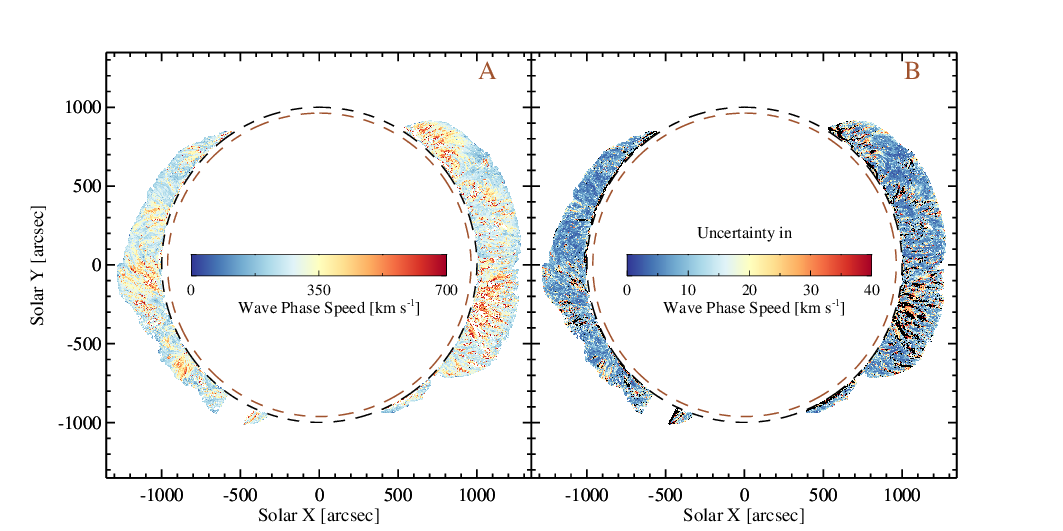}{\textwidth}{}}
    \caption{Phase speed determination: (A and B) Maps of the derived phase speed and the associated uncertainty. The circles are as in Figure $\ref{fig:doppler velocity and magnetic field}$.}
    \label{fig:wave_speed}
\end{figure*}
\begin{figure*}[t]
    \centering
    \gridline{\fig{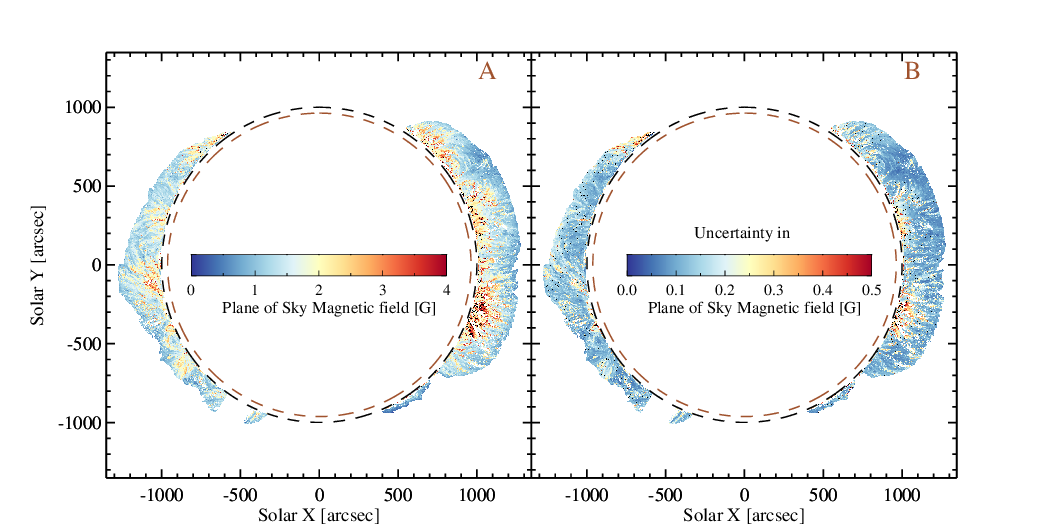}{\textwidth}{}}
    \caption{Coronal magnetic field: (A and B) Maps of plane-of-sky component of the coronal magnetic field and the associated uncertainty. The circles are as in Figure $\ref{fig:doppler velocity and magnetic field}$. This coronal magnetic field map corresponds to the magnetic field obtained using simple inversion. The number of pixels in this map, after applying all the filters used in evaluating the phase speed, is 40,306.}
    \label{fig:simple_magnetic}
\end{figure*}

\section{Instrument and Data\label{sec:Instrument}}
The Coronal Multichannel Polarimeter \citep[CoMP:][]{2008SoPh..247..411T} is a 20 cm aperture coronagraph mounted at Mauna Loa Solar Observatory in Hawaii and can perform spectroscopic observations at infrared wavelengths. Its spatial sampling is $\sim$\SI{4.35}{\arcsecond}, and its field-of-view (FOV) is about 1.05-1.35 $R_\odot$ from the solar center. This instrument measures the complete polarization state of the Fe \rom{13} coronal emission lines at 1074.7 and 1079.8 nm and He \rom{1} chromospheric line at 1083.0 nm. CoMP observations have measured periodic Doppler velocity disturbances that have been interpreted as the presence of transverse propagating MHD waves in the corona \citep{doi:10.1126/science.1143304,2009ApJ...697.1384T,2015NatCo...6.7813M,2019NatAs...3..223M}. \par
\cite{2007Sci...318.1574D,2009A&A...503..213G,Tian_2012,2017A&A...603A.101L} identified these waves as kink waves, which have Alfv{\'e}nic nature. Doppler velocity dataset is used to determine the phase speed of kink waves ubiquitously present in the solar corona \citep{doi:10.1126/science.1143304, 10.1088/0004-637x/697/2/1384,2015NatCo...6.7813M, 2016ApJ...828...89M}. The Doppler velocity dataset is taken from 20:39 UT to 21:26 UT on 2016 October 14. (Animation of Doppler velocity data is also available.) It should be noted that no flare or Coronal Mass Ejection (CME) happened during this time. Different structures present during the observations are shown in Appendix \ref{AppendixC}.

\section{Magnetic field estimation using Simple Inversion \label{sec:simple}}
To determine the phase speed, we consider 96 frames corresponding to  Fe \rom{13} 1074.7 nm from 20:39 UT to 21:26 UT on 2016 October 14. A constant time cadence of $\sim$30 s is ensured between these frames. The dataset obtained is usually associated with a certain degree of uncertainty, i.e. noise. Dataset is first aligned to remove the noise. Then a Gaussian filter corresponding to 3.5 mHz (corresponding to 5-minute oscillations) \citep{doi:10.1126/science.1143304,2009ApJ...697.1384T,2015NatCo...6.7813M,2019NatAs...3..223M} is applied to this Doppler velocity data. A filtered Doppler velocity dataset is obtained as shown in Figure \ref{fig:doppler velocity and magnetic field}(A). Furthermore, to reduce the effect of low signal-to-noise ratio in the data, only those pixels are used which have a peak intensity of Fe \rom{13} 1074.7 nm higher than 1ppm. The magnetic field in the solar corona acts as the waveguide as these transverse waves propagate along the magnetic field. Hence, the wave propagation direction is also required. Wave propagation direction is obtained using a cross-correlation technique as explained in \cite{2020Sci...369..694Y}. Only those pixels which have a cross-correlation value higher than 0.5 are considered. The wave propagation direction map is shown in Figure \ref{fig:doppler velocity and magnetic field}(B). The wave propagation direction is determined with respect to east-west orientation. However, an inherent 180$^{\circ}$ ambiguity limits the wave propagation direction map to a range between -90$^{\circ}$ and +90$^{\circ}$. The abrupt shifts in wave propagation angles at specific pixels are a consequence of this 180$^{\circ}$ ambiguity, where wave angles near -90$^{\circ}$ and +90$^{\circ}$ essentially indicate the same wave propagation directions. \par
After obtaining the wave propagation direction, the phase speed of the prevailing transverse waves in the FOV of CoMP is determined using the following procedure. First, we define the track of 31 pixels along the wave propagation direction. Following that, a space-time diagram is constructed for each pixel on the track, and then a 2D Fast-Fourier transform (FFT) is applied to obtain the $k$-$\omega$ diagram. The $k$-$\omega$ diagram separates the outward and inward propagating waves. The inverse FFT corresponding to the positive and negative frequency of $k$-$\omega$ diagram is computed to obtain the space-time diagrams corresponding to outward and inward propagating waves. These space-time diagrams are used to cross-correlate the time series at each pixel with the rest of the pixels of the same track. The slope of these distance-time plots gives the phase speed of the waves at each pixel. In Figure \ref{fig:wave_speed}, only those pixels with a phase speed less than 700 km s$^{-1}$ are shown. A few pixels also have a phase speed of more than 700 km s$^{-1}$. These pixels are near the occulting disc or at the outer edge of the FOV and are thus generally associated with large uncertainty. \par
% \citep{2015NatCo...6.7813M}.\par
To perform the coronal seismology, density estimates are required. In this study, we use the Fe \rom{13} line pairs and CHIANTI \citep{1997A&AS..125..149D} database. The intensity ratio of Fe \rom{13} lines at 1079.8 nm and 1074.7 nm is obtained using CoMP data from 19:24 UT to 20:17 UT corresponding to different heights in the CoMP FOV. This intensity ratio obtained from CoMP is then compared with the ratio of nearly the same wavelength range of Fe \rom{13} ion from the CHIANTI database. This period is chosen because it is similar in time to the wave tracking period. CHIANTI Version 9.0 \citep{2019ApJS..241...22D} is used. Then, from the calibration curves obtained using the CHIANTI database, electron density in the solar corona can be obtained as in Figure \ref{fig:density and its error}.\par
The phase speed of the transverse propagating kink waves and density data are used to estimate the magnetic field strength in the solar corona, first by simple inversion and then by Bayesian inference. The phase speed of the kink waves, $c_k$, can be expressed as,
\begin{equation}
    c_k^2 = \frac{{B_i}^2+{B_o}^2}{\mu_o(\rho_i+\rho_o)},
\end{equation}
 where B is the magnetic field strength, $\rho$ is the mass density, $\mu_o$ is the magnetic permeability of the free space, and the subscripts $i$ and $o$ represent the corresponding physical parameters inside and outside the coronal structures, respectively. In coronal plasma, magnetic pressure will dominate the thermal pressure because of the low plasma $\beta$ value, so $B_i \sim B_o$. Moreover, individual flux tubes are likely to be unresolved at the spatial resolution of CoMP, allowing us to take the density averaged inside and outside these flux tubes $(\langle \rho \rangle)$ within each spatial pixel and estimate the magnetic field strength via,
\begin{equation}
 \label{sec:simple_formula}
    c_k= \frac{B}{\sqrt{\mu_o \langle \rho \rangle}}.
\end{equation}
Substituting density data $(\langle \rho \rangle)$ and phase speed data $c_k$ at each pixel, the plane of sky component of the magnetic field is estimated using Equation \ref{sec:simple_formula}.
The magnetic field strength, obtained through a simple inversion process, is presented in Figure \ref{fig:simple_magnetic}. The estimated magnetic field strength primarily ranges between 1 and 4 G, with a corresponding uncertainty ranging from 0.1 to 0.5 G. \par
Also, it should be noted that for wave tracking, observations are taken from 20:39 UT to 21:26 UT, and for density estimations, observations are taken from 19:24 UT to 20:17 UT. Due to the dynamic nature of the solar corona, the measurement of physical parameters, such as coronal density, is subject to both statistical and systematic uncertainties. To improve the magnetic field strength estimation, it is recommended to use an inversion process that considers the uncertainties in the measurements of physical parameters, such as coronal density. Bayesian inference provides a suitable approach for this, as it incorporates the uncertainty in density when determining the range of plausible magnetic field values.

\section{Magnetic field estimation using Bayesian Inference\label{sec:Bayesian}}
Bayesian analysis enables to perform parameter inference, model comparison, and model averaging applications to gain information about the magnetic field in the solar corona. This approach to parameter inference is based on the use of Bayes Theorem \citep{1763RSPT...53..370B}, 
\begin{equation}
\label{first_eq}
p(\theta|d,M) = \frac{p(d|\theta, M)p(\theta|M)}{\int p(d|\theta, M)p(\theta|M)d\theta}.   
\end{equation}
In this expression, $p(\theta|M)$ is the prior probability density for the parameter vector $\theta$, and $p(d|\theta, M)$ is the likelihood function. Both quantities are conditional on the assumed model $\textit{M}$. Their combination leads to the posterior probability distribution $p(\theta|d, M)$, which contains all the information that can be inferred from the observed data and the assumed model.\par
It should be noted that Bayesian inference does not provide point estimates for parameters but rather the full posterior probability distribution function (PDF), $p(\theta|d, M)$, which captures the entire range of plausible values for the parameters. This is a significant advantage over simple inversion.\par
In parameter inference, a parameterized model may have multiple parameters, $\theta = {\theta_1,...,\theta_i,...,\theta_N}$. The marginal posterior distribution for a parameter of interest, $\theta_i$, can be obtained by integrating the full posterior over the remaining nuisance parameters.
\begin{equation}\label{basic_marginal}
    p(\theta_i|d) = \int p(\theta|d)d\theta_1...d\theta_{i-1},d\theta_{i+1},...,d\theta_N.
\end{equation}
This results in the marginal posterior distribution for model parameter $\theta_i$, which includes all the information available in the prior knowledge and observed data. The uncertainty from the nuisance parameters will also propagate correctly to the one of interest.\par
Our model $\textit{M}$ is given by the theoretical prediction for the kink speed in the extended wave-length approximation, given by Equation (\ref{sec:simple_formula}). This is a two-parameter function, hence $\theta = \{\rho, B\}$ in our Bayesian inference model. To estimate the probability of the magnetic field strength using Bayesian inference, we first adopt prior distributions for the two model parameters, $\rho$, $B$. For the plasma density, we use measured density values and associated uncertainties from \cite{2020Sci...369..694Y} instead of synthetic density values as used by \cite{2011ApJ...740...44A}. Two different prior distributions are used: a Gaussian prior and a uniform prior. The uniform prior assigns an equal probability for the density in the range $[\mu_{\rho_i} - 3\sigma_{\rho_i},\mu_{\rho_i} + 3\sigma_{\rho_i}]$, where ${\mu_{\rho_i}}$ represents the measured density value and ${\sigma_{\rho_i}}$ represents the uncertainty for the $i^{th}$ pixel. The uniform prior probability for density is described by the following equation:
\begin{equation}\label{uniform_prior_density}
    p(\rho) =\begin{cases}
    \frac{1}{(\mu_{\rho_i}+3\sigma_{\rho_i}) - (\mu_{\rho_i}-3\sigma_{\rho_i})} & \text{$\mu_{\rho_i}-3\sigma_{\rho_i} \le \rho\le \mu_{\rho_i}+3\sigma_{\rho_i}$} \\ 
    0 & \text{otherwise}.
    \end{cases}
\end{equation}
The Gaussian prior for the density at each pixel in the FOV is given by:
\begin{equation}\label{Gaussian_prior}
    p(\rho) = \frac{1}{\sqrt{2\pi}\sigma_{\rho_i}}  \exp{\frac{-(\rho - \mu_{\rho_i})^2}{2\sigma_{\rho_i}^2}}.
\end{equation}
For the magnetic field strength, we restrict the possible range of variation to the one obtained using simple inversion, i.e., [$B_{\textit{min}}$, $B_{\textit{max}}$] = [1,8] G, with $B_{\textit{max}}$ only present at a few pixels. For the magnetic field, we restrict the possible range of variation to the one obtained using simple inversion. Within this range, we assign a uniform prior probability for the magnetic field strength described by the following equation:
\begin{equation}\label{prior_mag}
    p(B) =\begin{cases}
    \frac{1}{B_{\textit{max}} - B_{\textit{min}}} & \text{$B_{\textit{min}} \le B\le B_{\textit{max}}$} \\ 
    0 & \text{otherwise}.
    \end{cases}
\end{equation}
For our computations, we constructed a two-dimensional grid over the parameter space with $N_{\rho}=10,000$ and $N_{B}=71 $ points, which covers the ranges of considered values for plasma density and magnetic field strength.\par
An additional magnetic field prior, known as the gamma prior, is also considered. The probability distribution function of the gamma prior is given by Equation \ref{gamma_prior_df}. Further details of how the parameters of this distribution are evaluated can be found in Appendix \ref{AppendixA}.\par
In the Bayesian framework, the data are fixed, and the model parameters are unknown to which a probability value can be assigned. The general expression of the likelihood function is the numerator of Equation (3), $p(d | \theta, M)$,  explicitly indicates that this function is conditional on the model \textit{M}, which is considered to be true.  It does not represent the likelihood of different data realisations.  Its purpose is to measure the discrepancy between model predictions and observed data as a function of the model parameters, taking as reference the uncertainty of the data, thus assigning different levels of likelihood to alternative parameter combinations. In our case, the phase velocity is forward modelled using model $M$ in Equation (\ref{first_eq}) for different combinations of the parameter values defined by those priors. These theoretical predictions $v_p$ are compared to the observed phase speed $c_k$ at each pixel, and the relative merit (likelihood) of each combination is assessed by the adoption of a Gaussian likelihood function of the form 
\begin{equation}\label{likelihood}
    p(c_k|B,\rho, M) = \frac{1}{\sqrt{2\pi}\sigma_{c_k}}  \exp{\frac{-(c_k - v_p(B, \rho))^2}{2\sigma_{c_k}^2}}.
\end{equation}
Here, $\sigma_{c_k}$ corresponds to the uncertainty in the phase speed obtained earlier. This uncertainty is different at each pixel and corresponds to the uncertainty obtained while evaluating the phase speed $c_k$. The joint posterior probability of the magnetic field and density in the FOV, given our model is true, is obtained as
\begin{equation}\label{posterior}
    p(B,\rho|c_k) = \frac{p(c_k|B,\rho, M)p(B)p(\rho)}{\int \int p(c_k|B,\rho,M)p(B)p(\rho)dB d\rho},
\end{equation}
where $p(c_k|B,\rho, M)$ is the likelihood function and $p(B)$, $p(\rho)$ are the prior probability distribution functions of the magnetic field and density, respectively. We have done forward modelling of phase speed with the measured phase speed and the corresponding error for the likelihood function. All the posteriors in this analysis are normalized.\par
\begin{figure*}
    \centering
    \includegraphics{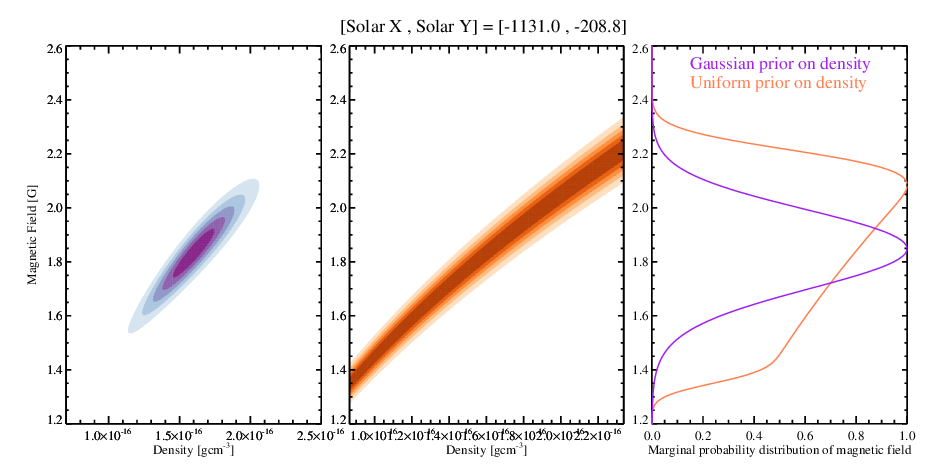}  
    \caption{Left and middle panels represent the joint probability distributions of magnetic field and density at the specified location. For Gaussian prior on density [Left panel], it is maximum for density value for which the prior probability is also maximum, but for a uniform prior on density. The right panel represents the marginal magnetic field obtained by integrating the joint probability distributions with density.}
    \label{fig:joint_probability}
\end{figure*}
\begin{figure*}[t!]
    \centering
    \includegraphics[height=0.3\textheight]{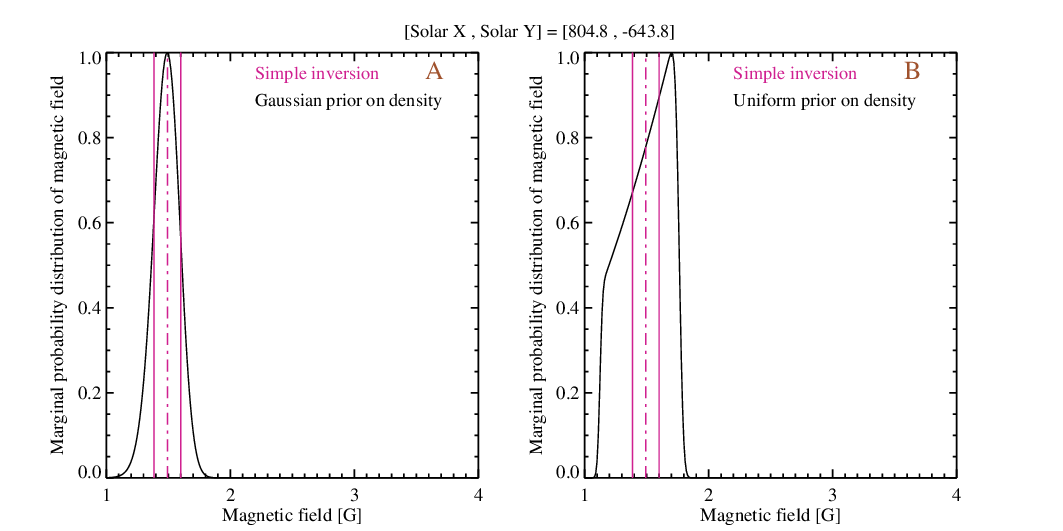}
    \includegraphics[height=0.3\textheight]{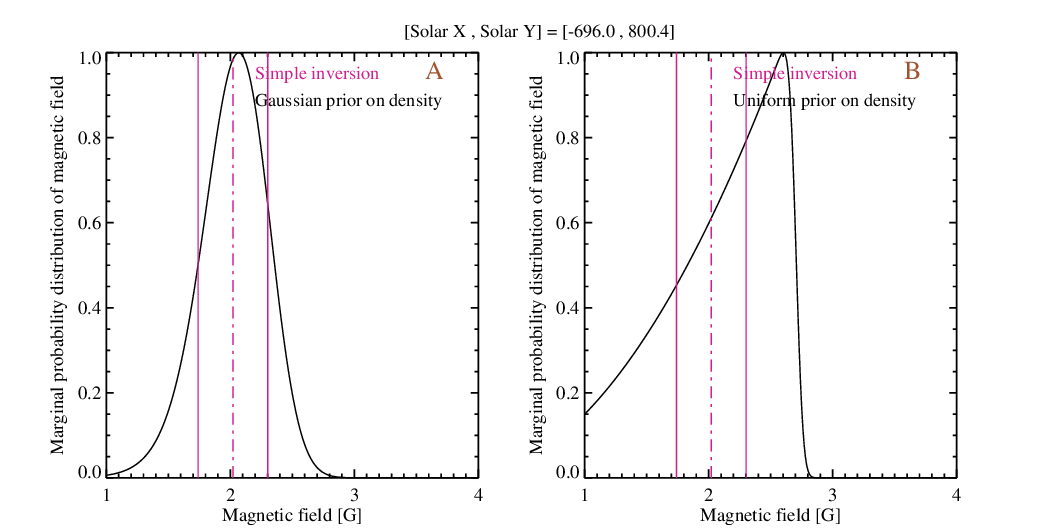}
    \caption{Marginal magnetic field: (A) Marginal magnetic field at two locations using Gaussian prior on density. (B) Marginal magnetic field at the same locations using a uniform prior on density. The vertical dot-dash violet line corresponds to the magnetic field at the same location using simple inversion, and the vertical solid violet line corresponds to the lower and upper limits of the magnetic field. Phase speed and associated uncertainty at location (804.8,-693.8) ((-696,800.4)) is  \SI{475(49)}{\kilo\meter\per\second}(\SI{275(4)}{\kilo\meter\per\second}), and density and associated uncertainty values are \SI{1.98(0.44)e-13}{\kg\per\cubic\meter}(\SI{4.20(1.16)e-13}{\kg\per\cubic\meter}). These results in the magnetic field and associated uncertainty obtained using simple inversion as \SI{2.4(0.4)}{G} (\SI{2.0(0.3)}{G}).}
    \label{fig:Marginal_at_each_pixel}
\end{figure*}
The relative simplicity of the forward and inverse problems (Equations \ref{sec:simple_formula} and \ref{posterior}), which involve a two-dimensional parameter space and a one-dimensional observable space, makes feasible the computation of our Bayesian posteriors using direct numerical evaluation and integration over a grid of points in parameter space. These are performed using routine IDL and Python scripts. \cite{2019A&A...625A..35A} have shown that this grid-approximation method and a Markov chain Monte Carlo (MCMC) sampling of the posterior give essentially the same results to the solution of kink wave phase speed inversion problem (see their Appendix A). In our study, we have verified the good correspondence between the direct integration and the MCMC solutions, described in Appendix \ref{AppendixB}.\par
Using a uniform prior on magnetic field (Equation \ref{prior_mag}) and two different priors on density (Equation \ref{uniform_prior_density} and \ref{Gaussian_prior}) and the likelihood function (Equation \ref{likelihood}), Equation \ref{posterior} helps evaluate the joint posterior probability distribution of the magnetic field and density. The left and middle panels of Figure \ref{fig:joint_probability} represent the joint probability distribution of the magnetic field and density obtained at one location using two different priors on density. The shape of the inferred joint probability distribution differs for different priors on density. The use of a Gaussian prior on density leads to a more constrained inference because it provides more information to the inference process. From the joint posteriors, our interest is in estimating the marginal probability distributions of the magnetic field strength; this is done by integrating the joint posterior probability distribution over the density as follows. This marginalized distribution of the magnetic field is determined using an equation similar to the Equation \ref{basic_marginal} as,
\begin{equation}\label{marginal_b}
    p(B|c_k) = \int p(B,\rho|c_k)d\rho.
\end{equation}
The evaluation of the marginal probability distributions of the magnetic field also incorporates the uncertainty in density as it involves integrating the whole joint posterior probability with respect to density. The rightmost panel of Figure \ref{fig:joint_probability} represents the marginal probability distribution of the magnetic field. For Gaussian prior on density, the marginal probability distribution of the magnetic field peaks where the joint probability distribution of the magnetic field and density is maximum. The magnetic field value for which the marginal probability distribution is maximum is the most probable value of the magnetic field for that location. \par
Figure \ref{fig:Marginal_at_each_pixel}(A) represents the marginal posterior for the magnetic field inferred at two other locations in the FOV, using a uniform prior on the magnetic field and two different priors on density. The vertical dash-dotted and solid violet lines indicate the magnetic field estimate and the associated uncertainty obtained using simple inversion. Our results show that the simple inversion point estimates and the Bayesian maximum a posteriori estimates do not always coincide. The magnitude of the discrepancy depends on the particular location and the type of prior on density employed in the inference. The difference is larger on location  [-696.0,800.4] (bottom panels) than on location [804.8, -643.8] (top panels). The reason for this is that the relative uncertainty on density at [-696.0,800.4] is 27\% while at location [804.8, -643.8] and [-1131.0, -208.8], the relative uncertainty in density is 22\% and 15\%. The mismatch between the simple and Bayesian estimates is larger when using a uniform prior on density (right panels) than when a Gaussian prior is employed (left panels). Here, again, the use of a Gaussian prior is helping in obtaining a more constrained inference result.\par
We have explored another option for the prior on magnetic field strength, with the use of a Gamma distribution. This has the advantage of being strictly positive and with no upper boundary. A summary of inference results obtained using this prior for the magnetic field strength is presented in Appendix A.
One comparison between the marginal magnetic field distribution evaluated using a uniform prior on the magnetic field and gamma prior on the magnetic field is also made at location [-1131, -208.8] in Figure \ref{fig:comparison_gamma_uniform_marginal}. The violet lines are as mentioned above. The royal blue curve corresponds to the marginal magnetic field distribution obtained using a uniform prior on the magnetic field, whereas the black curve corresponds to the gamma prior on the magnetic field for both density distributions. The marginal magnetic field distribution at the other two locations is shown in Figure \ref{fig:Gamma_Marginal_at_each_pixel}. The area under the plots of marginal magnetic field versus magnetic field will be unity, for the pixels having higher values of the marginal distribution of the magnetic field will result in giving the more constrained plausible range of magnetic field and vice versa. This can be seen from Figure \ref{fig:comparison_gamma_uniform_marginal} for the uniform prior on the magnetic field, which has a more constrained curve as compared to the other locations as in Figure \ref{fig:comparison_gamma_uniform_marginal}. Furthermore, one can easily check that the magnetic field distribution obtained using gamma prior on the magnetic field gives more constrained results than that of a uniform prior on the magnetic field. 
\par

\begin{figure}[h!]
    \centering
    \includegraphics[height = 6.5cm]{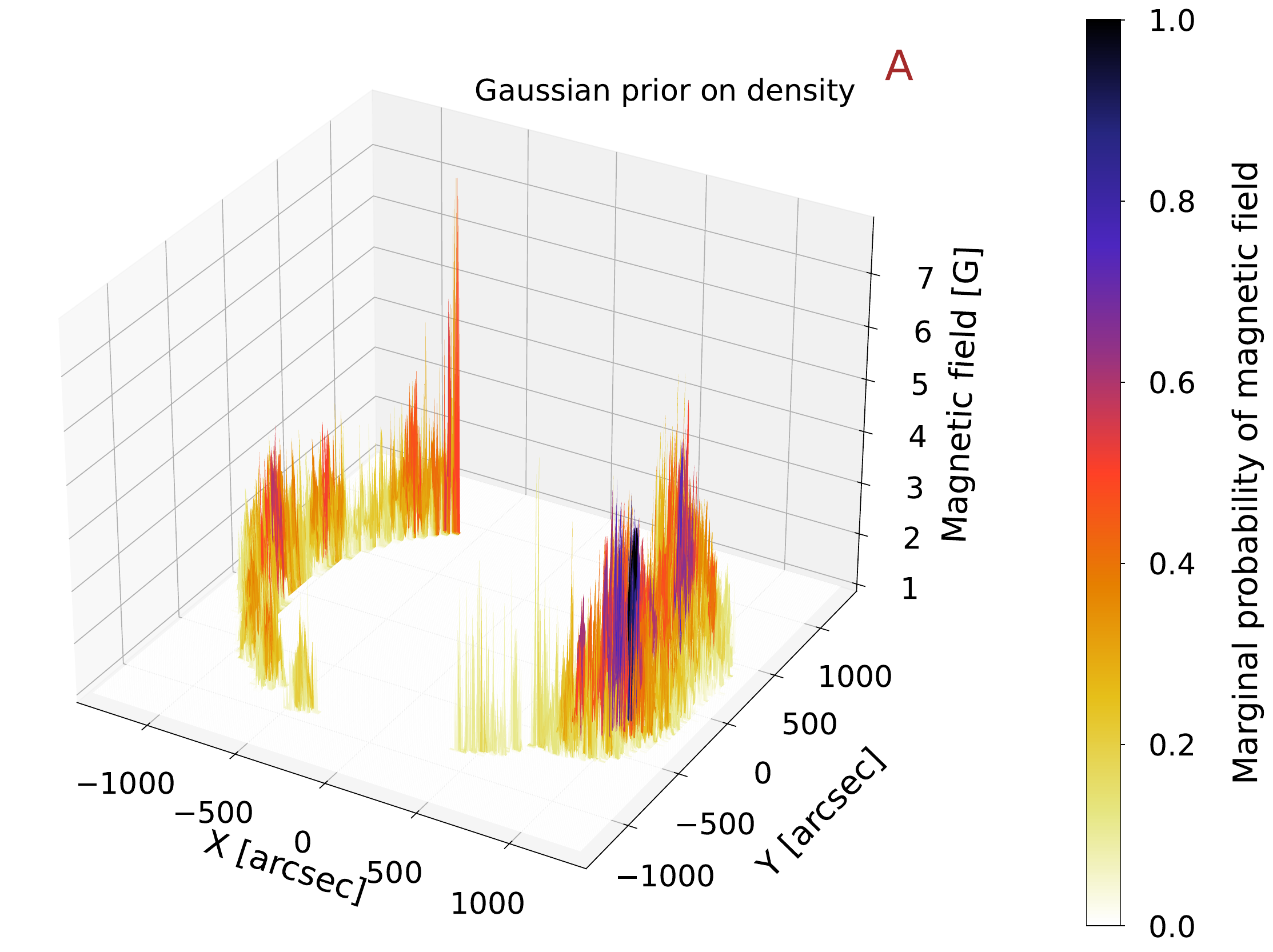}
    \includegraphics[ height = 6.5cm]{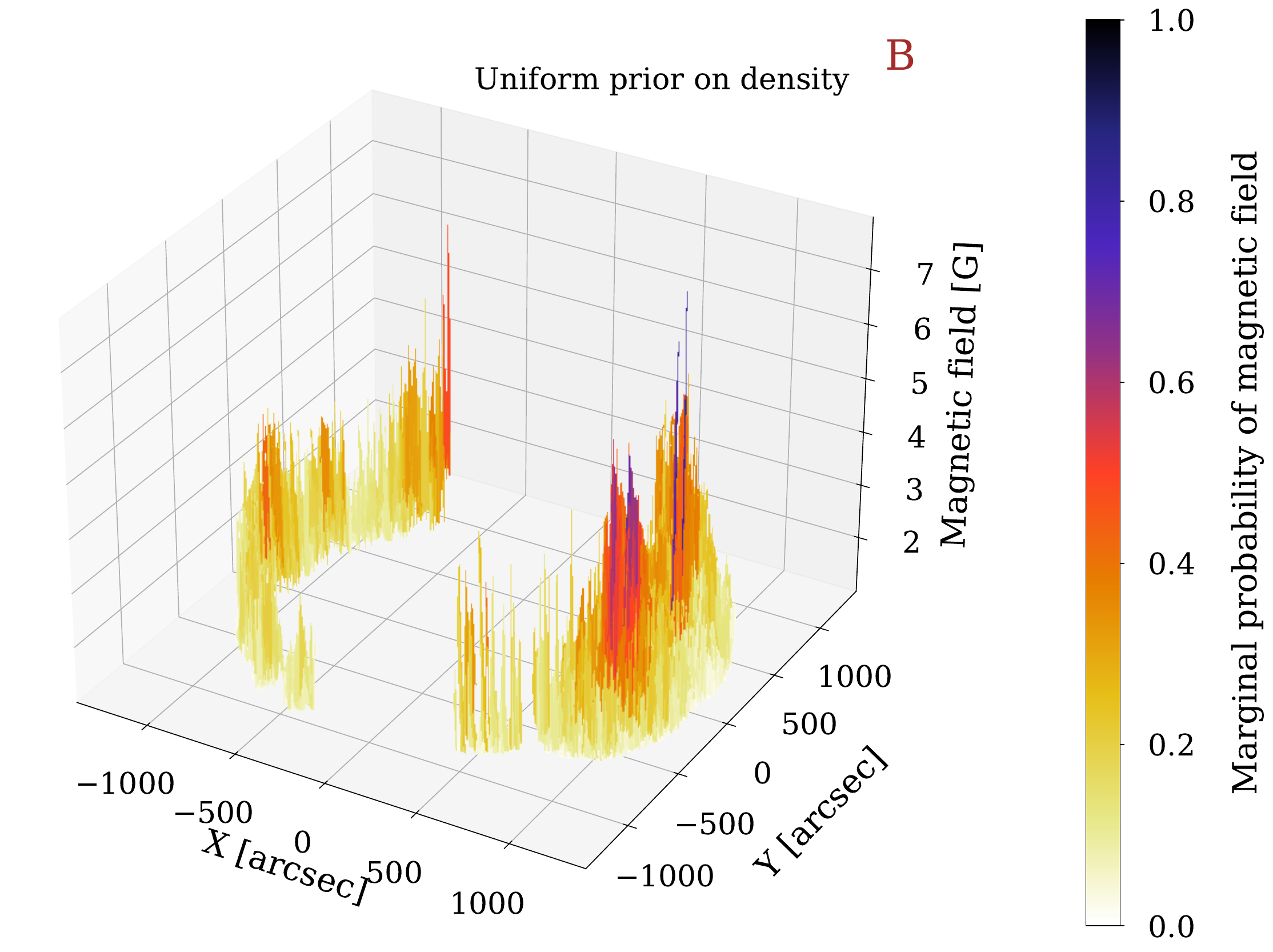}
    \caption{[A] Marginal distribution of magnetic field obtained using Gaussian prior density distribution and a uniform prior magnetic field distribution in the whole FOV of CoMP. [B] Marginal magnetic field obtained using a uniform prior density distribution and a uniform prior magnetic field distribution in the whole FOV.}
    \label{fig:marginal_magnetic_field}
\end{figure}

Further, we computed the marginal probability distribution of the magnetic field in the whole FOV of CoMP. For this, we normalise the marginal probability of the magnetic field at each location with respect to the maximum value of the marginal probability obtained in the whole FOV.  The results are shown in Figure \ref{fig:marginal_magnetic_field}. Here, the height of each bar represents the range of plausible values of the magnetic field for that location, and the colour bar represents the colour corresponding to the marginal probability of the most probable magnetic field at each location. The larger the height of the bar, the more the values of the magnetic field are plausible. Similar maps using gamma prior on the magnetic field are shown in $\ref{fig:marginal_gamma_magnetic_field}$ in Appendix (\ref{AppendixA}). Further, the convergence of the magnetic field is also shown by using Markov chain Monte Carlo (refer Appendix \ref{AppendixB})\par
\color{black}
\begin{figure*}[b!]
    \centering
    \includegraphics[width=\textwidth,height=9cm]{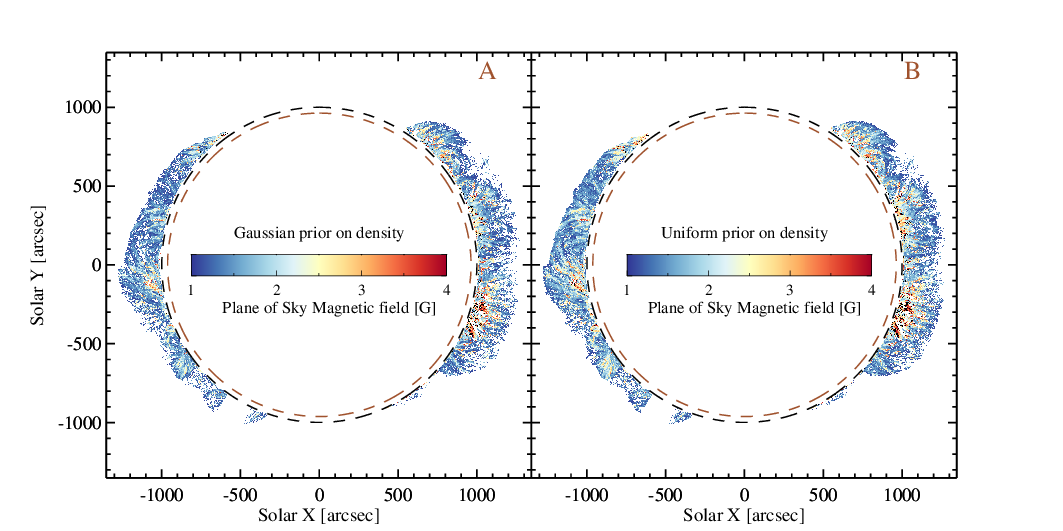}
% \gridline{\fig{image_6.eps}{\textwidth}{}}
    \caption{Coronal magnetic field using Bayesian Inference: (A) Map of plane-of-sky component of the coronal magnetic field obtained using Gaussian prior on density. (B) Map of the plane of sky component of the magnetic field obtained using a uniform prior on density. The circles are as in Figure $\ref{fig:doppler velocity and magnetic field}$. }
    \label{fig:bayesian_magnetic}
\end{figure*}
\begin{figure*}[t!]
    \centering
    \gridline{\fig{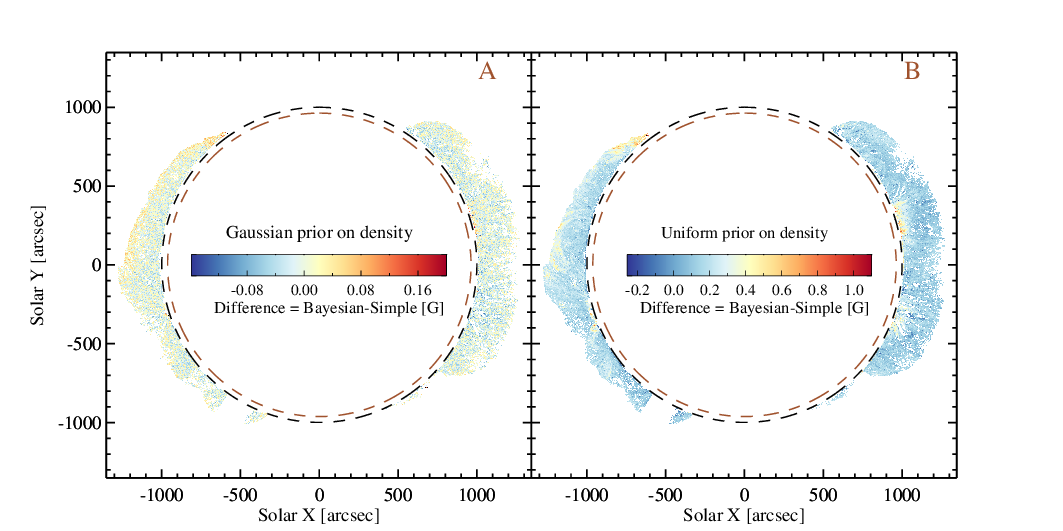}{\textwidth}{}}
    \caption{Magnetic field obtained using Bayesian inference versus simple inversion: (A) Difference between the plane-of-sky component of magnetic field obtained using Bayesian inference and simple inversion at each pixel in CoMP FOV using Gaussian prior on density. (B) Difference between the plane-of-sky component of magnetic field obtained using Bayesian inference and simple inversion at the same pixel in CoMP FOV using a uniform prior on density.}
    \label{fig:difference}
\end{figure*}
The simple inversion approach yields a single-valued magnitude of the magnetic field at each pixel, along with its associated uncertainty. On the other hand, Bayesian inference produces a probability distribution corresponding to each magnetic field value at each pixel. Figure \ref{fig:bayesian_magnetic} shows the Bayesian analogue of the map in Figure \ref{fig:simple_magnetic}A, representing now the maximum a posteriori estimates for the magnetic field strength, for the two considered priors in density. As explained above, the inference results have a dependence on the prior assumption on density, because the uniform prior is less informative that the Gaussian prior. These differences can be appreciated at several pixels in the inference results displayed in Figure \ref{fig:bayesian_magnetic}.\par
The difference between the magnetic field obtained through simple inversion and the most probable magnetic field obtained through Bayesian inference is illustrated in Figure \ref{fig:difference}. In most regions, the difference is close to zero. However, the difference is more pronounced near the occulter or at the edge of the FOV, where the uncertainty in density or phase speed is large. It should be noted that the difference is larger for the uniform prior on density as compared to the Gaussian prior on density. The most probable magnetic field obtained using Gaussian prior on density is not very much different from the magnetic field estimate obtained using simple inversion.\par
\begin{figure*}
    \centering
    \includegraphics[width =0.47\textwidth]{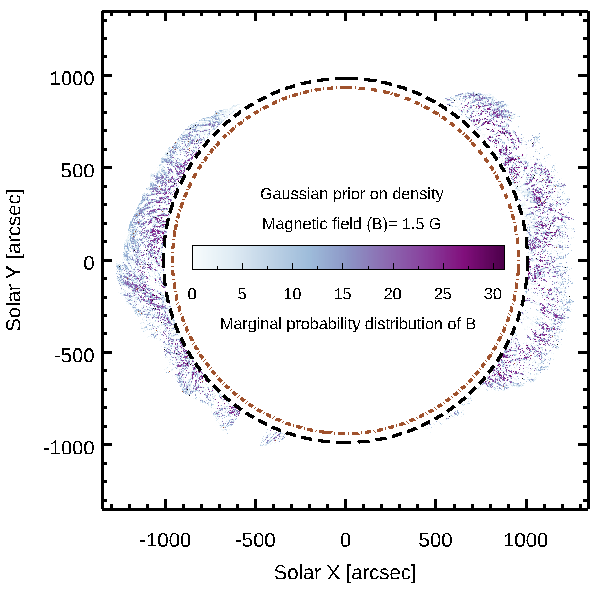}
    \includegraphics[width =0.47\textwidth]{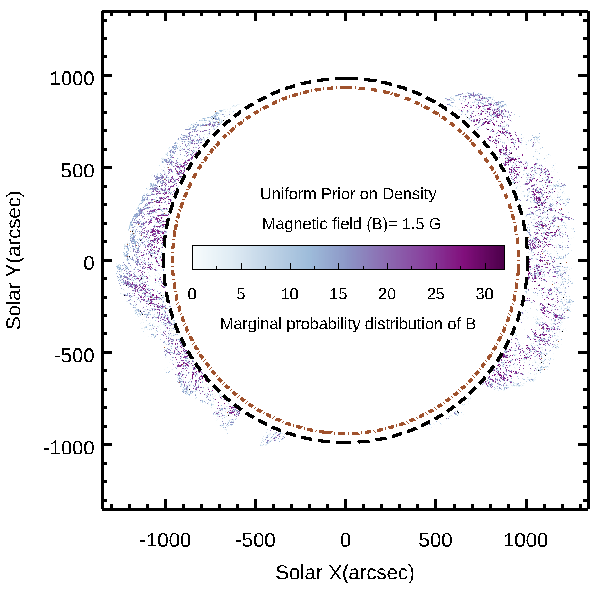}
    \caption{Variation of marginal probability with magnetic field: Left and Right panel represent the different marginal probability values of the magnetic field (= 1.5 G) at different pixels for Gaussian prior on density and uniform prior on density, respectively. An animation of the variation of marginal probability values with the variation of the magnetic field is available. The animation shows the changes from a magnetic field of 1.0 G to 8.0 G in increments of 0.1 G. The animation shows that the lower magnetic field values are more probable in the outer FOV, while higher magnetic field values are more probable near the inner edge of the FOV. The animation has a real-time duration of 5 seconds. }
    \label{fig:marginal_field_versus_magnetic_field}
\end{figure*}

The Bayesian inference results enable us to compare the probability of a given value for the magnetic field strength at different locations in the FOV. This is an advantage with respect to the simple inversion result, with which this comparison cannot be drawn. Figure \ref{fig:marginal_field_versus_magnetic_field} represents the marginal probability of magnetic field = 1.5 G at all the pixels in the field of view of CoMP. For 1.5 G, the magnetic field is more structured and concentrated at the flux tubes. Lower values of magnetic fields are most probable at the outer edges of FOV, whereas higher values of the magnetic field are most probable near the inner edge of the FOV. One can also conclude from the movies that the higher values of the magnetic field are most probable in the west limb. From the animations, it is clear that the marginal distribution of the magnetic field, when plotted in the whole FOV by varying the strength of the magnetic field, gives the region where those magnetic field values are most probable. \par
A similar analysis was also performed with a different range of density in uniform prior where density range is $[\mu_{\rho_i} - \sigma_{\rho_i},\mu_{\rho_i} + \sigma_{\rho_i}]$. With this constrained prior in density, the marginal distribution of the magnetic field was also constrained. The most probable magnetic field obtained using this density range was similar to the one obtained using Gaussian prior on density.
In Section \ref{sec:simple_formula}, it was noted that phase speeds exceeding 700 km s$^{-1}$ were observed in a few pixels. To evaluate the impact of these observations, the analysis was extended to include phase speeds up to 2000 km s$^{-1}$, resulting in a revised maximum magnetic field strength of 21 G instead of 8 G as obtained through simple inversion. This new magnetic field range, B $\epsilon$ [1,21] G, was then considered to estimate the magnetic field strength using Bayesian inference. The pixels having phase speed less than 700 km s$^{-1}$ were found to yield the same results. The same analysis was repeated for a different range of magnetic field, B $\epsilon$ [1,25] G, to check if it has any impact; however, again, similar results were obtained.\par
\subsection{Radial variation of the coronal magnetic field}
\begin{figure*}
    \centering
    \includegraphics[trim={0 0 7cm 0},clip]{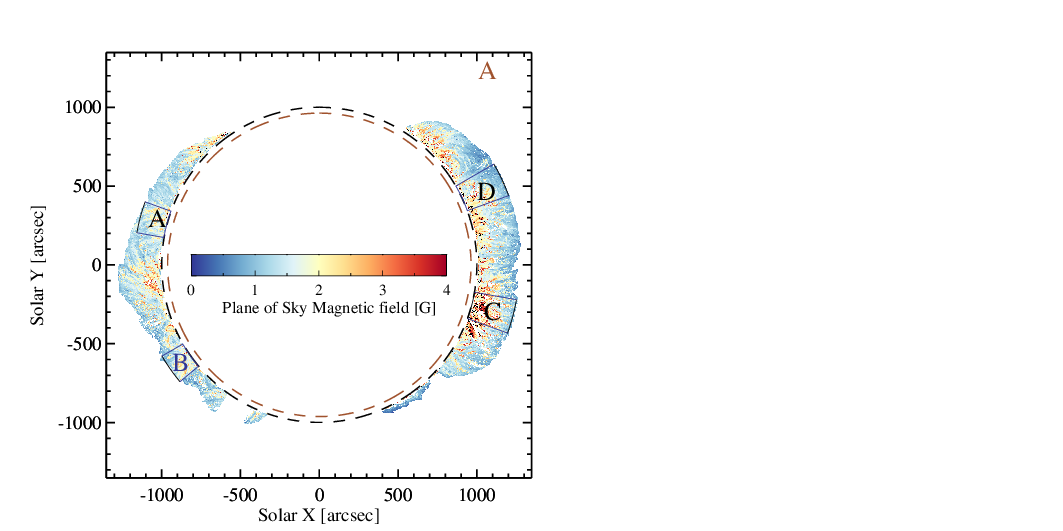}
    % \hspace{0.4cm}
    \includegraphics[width=0.40\textwidth]{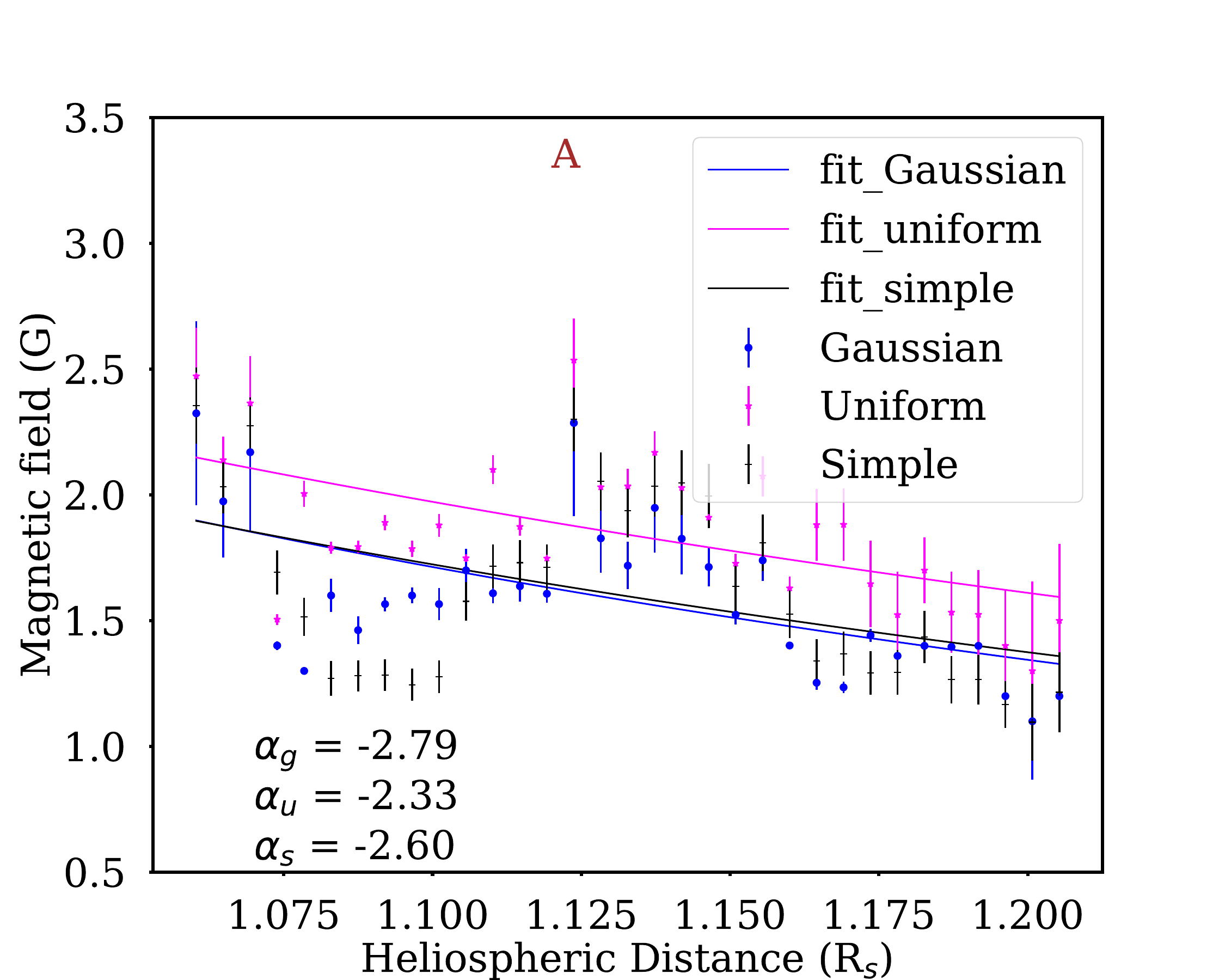}
    \includegraphics[width=0.40\textwidth]{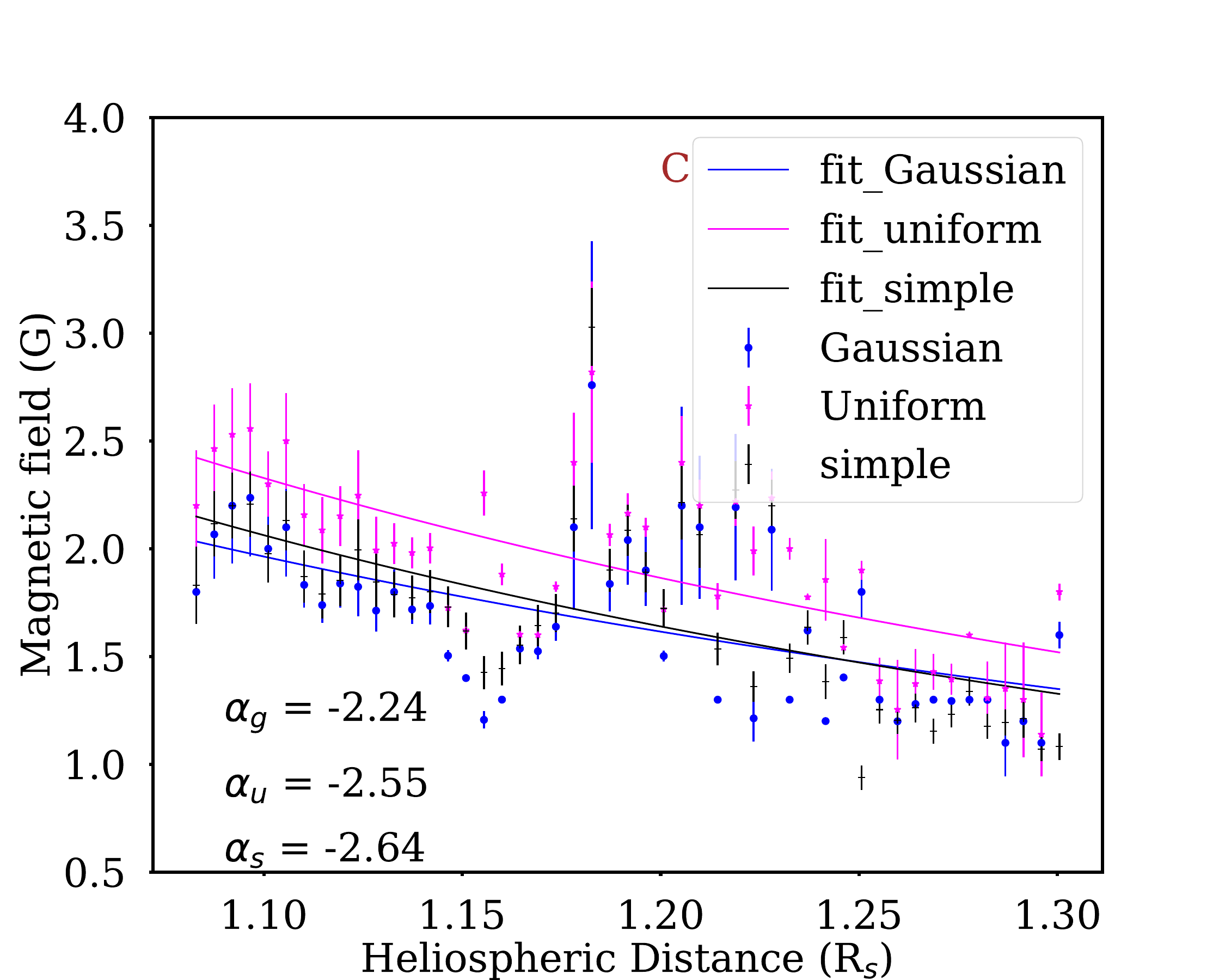}
    \includegraphics[width=0.40\textwidth]{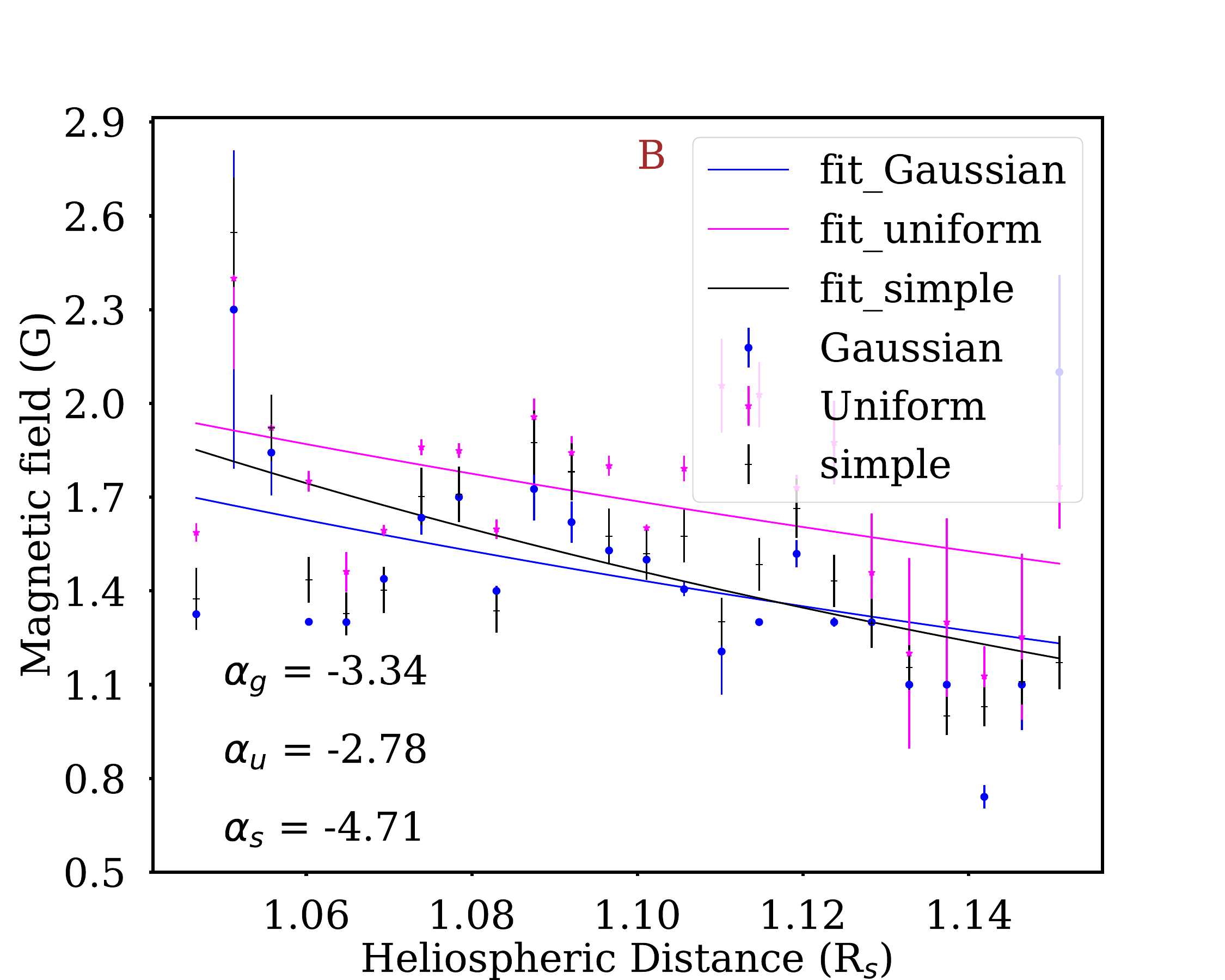}
    \includegraphics[width=0.40\textwidth]{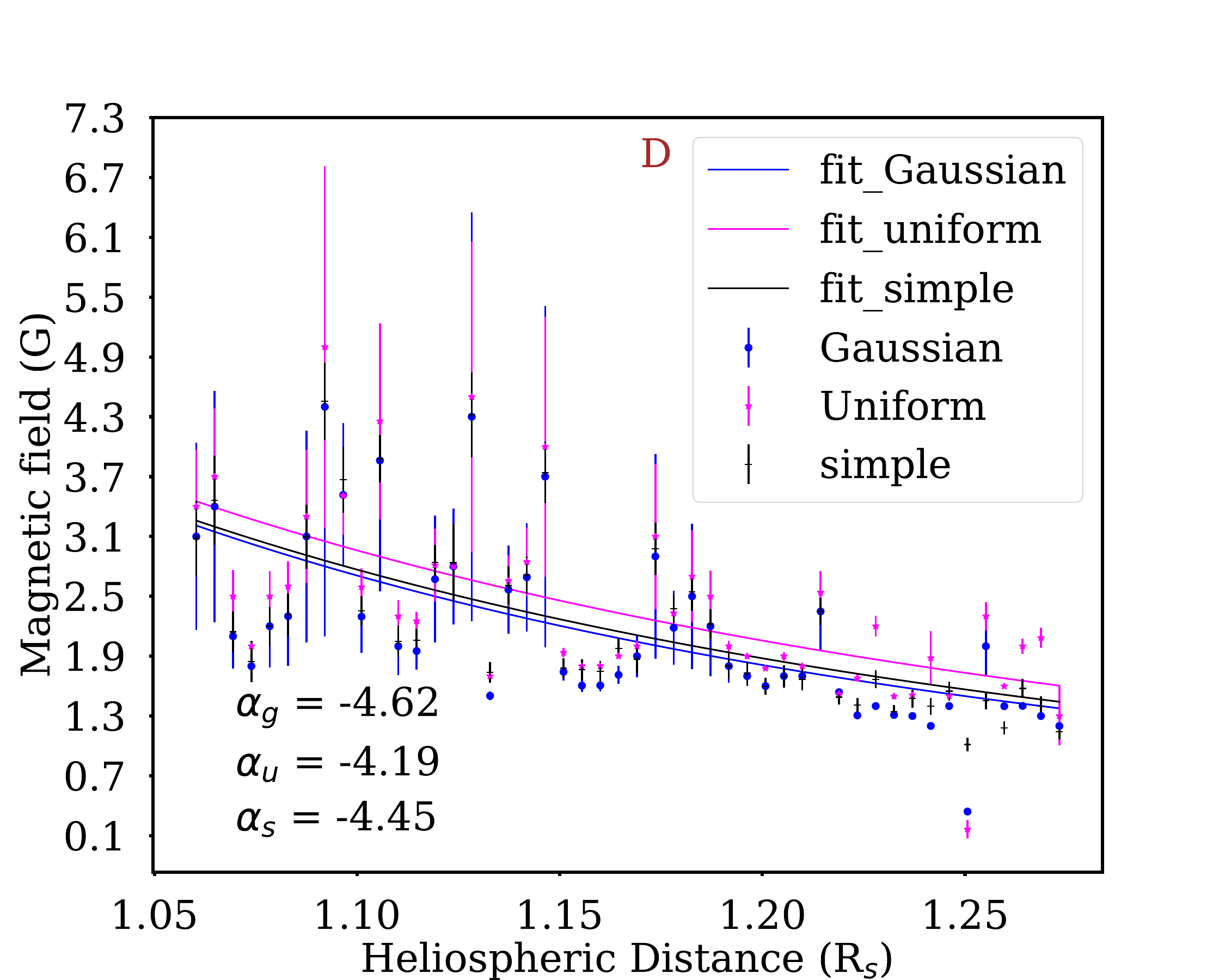}
    \caption{ Radial profile of magnetic field: Top panel shows the plane of sky component of the magnetic field (same as in Figure \ref{fig:bayesian_magnetic}) with four regions marked in the east and west limb. (A), (B), (C) and (D) show the variation of average coronal magnetic field strength as a function of radial distance from the solar centre for the marked regions.} 
    \label{fig:radial_profile}
\end{figure*}

The radial variation of the coronal magnetic field obtained using simple inversion and Bayesian inference is shown in Figure \ref{fig:radial_profile}. The region on the east limb represents the quiet Sun (QS) region, whereas the region on the west limb represents the active region (AR). Previous investigations \citep{2019ApJ...881...24K,2020ScChE..63.2357Y} have shown that the radial variation of the coronal magnetic field follows a power-law function. The power-law function ($Ar^{\alpha}$) is fitted to each curve as shown in Figure \ref{fig:radial_profile}, and the power-law index ($\alpha$) is obtained.  The power law index value for each prior on density distribution in Bayesian inference and simple inversion is almost the same.\par
\subsection{Magnetic field along and across the coronal loops}
\begin{figure*}
    \centering
        \begin{center}
          \includegraphics[height =0.35\textheight]{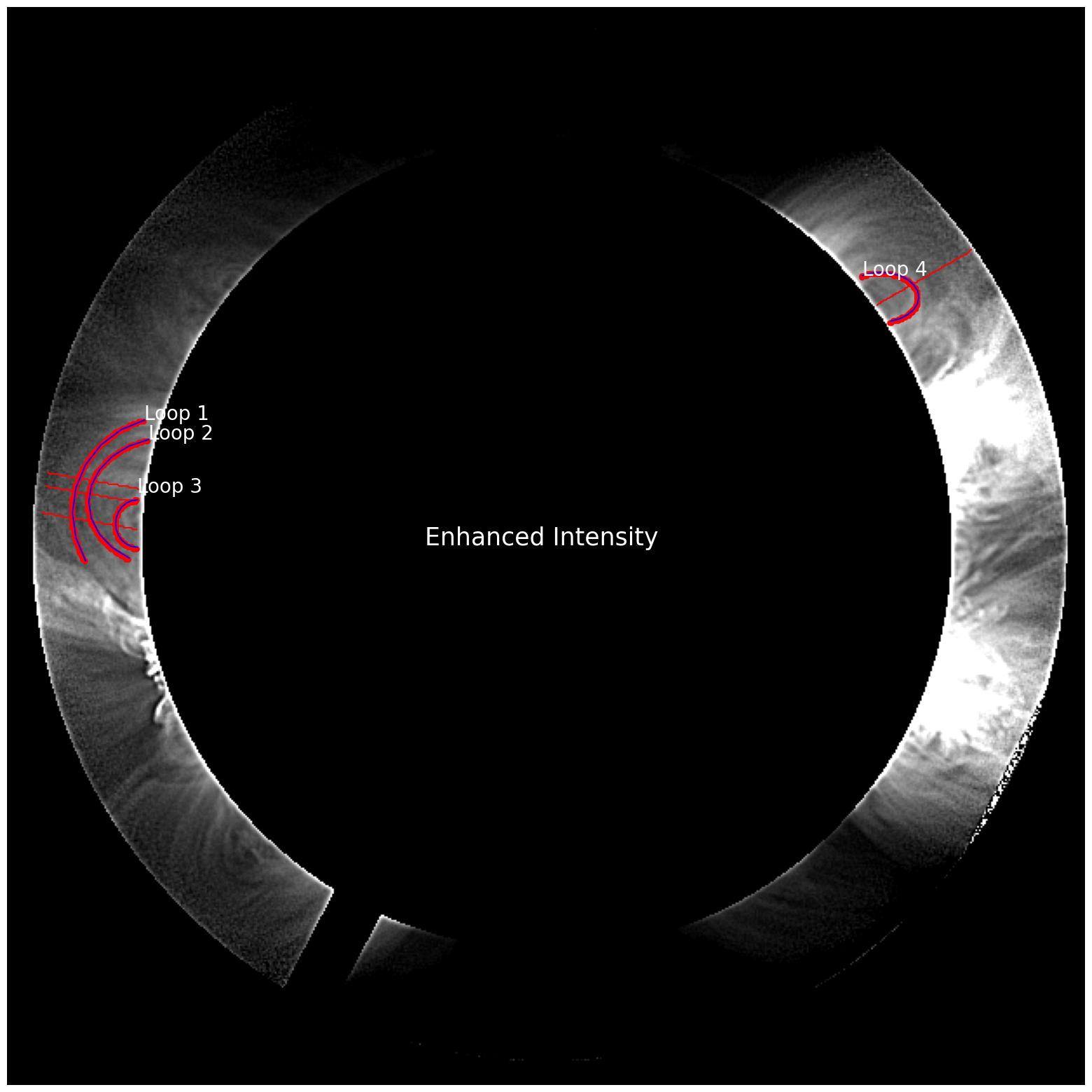}  
        \end{center}
        \includegraphics[width = 0.23\textwidth]{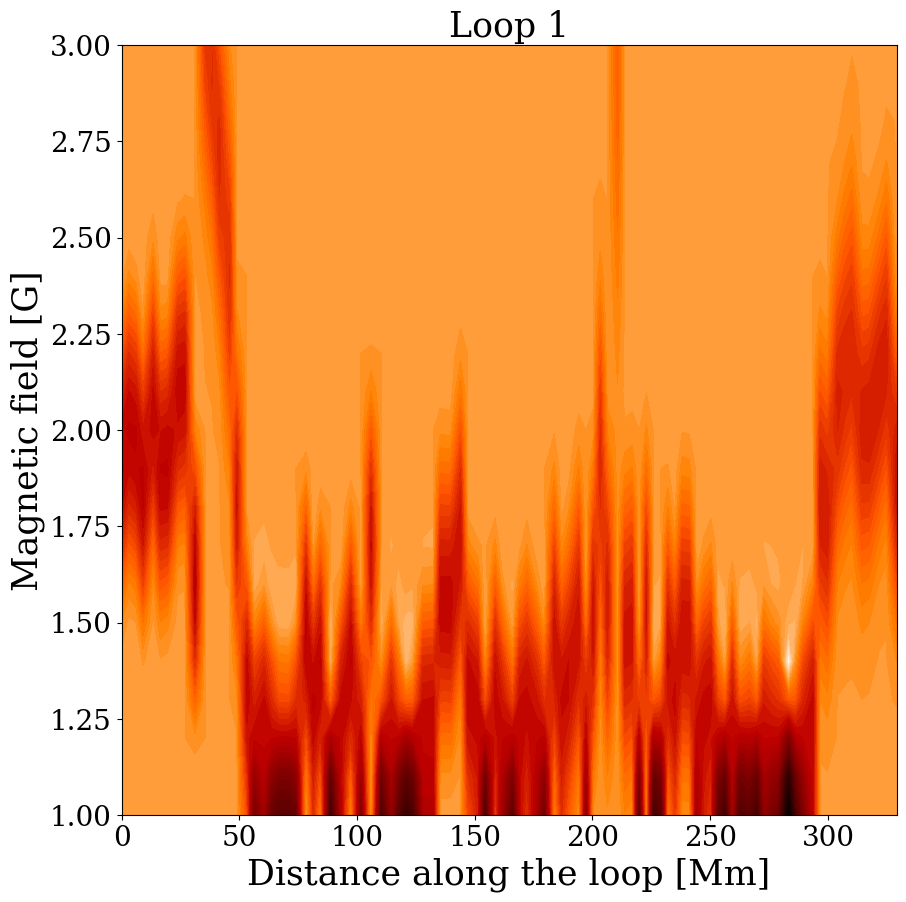}
        \includegraphics[width =0.23\textwidth]{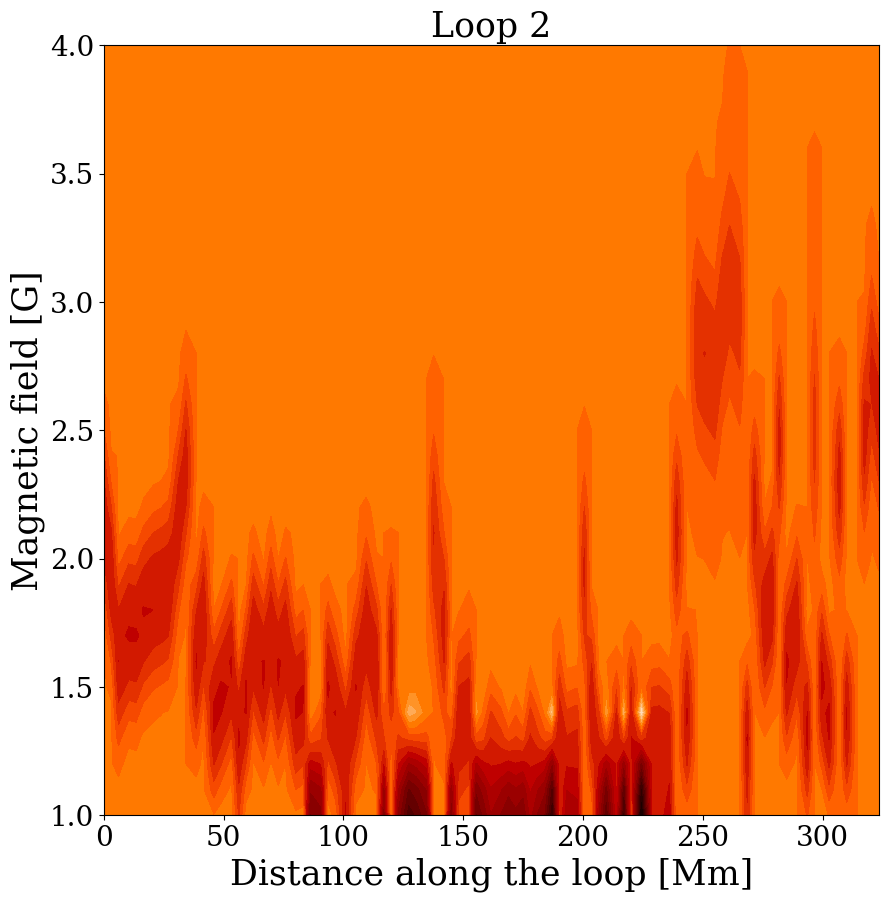}
        \includegraphics[width = 0.23\textwidth]{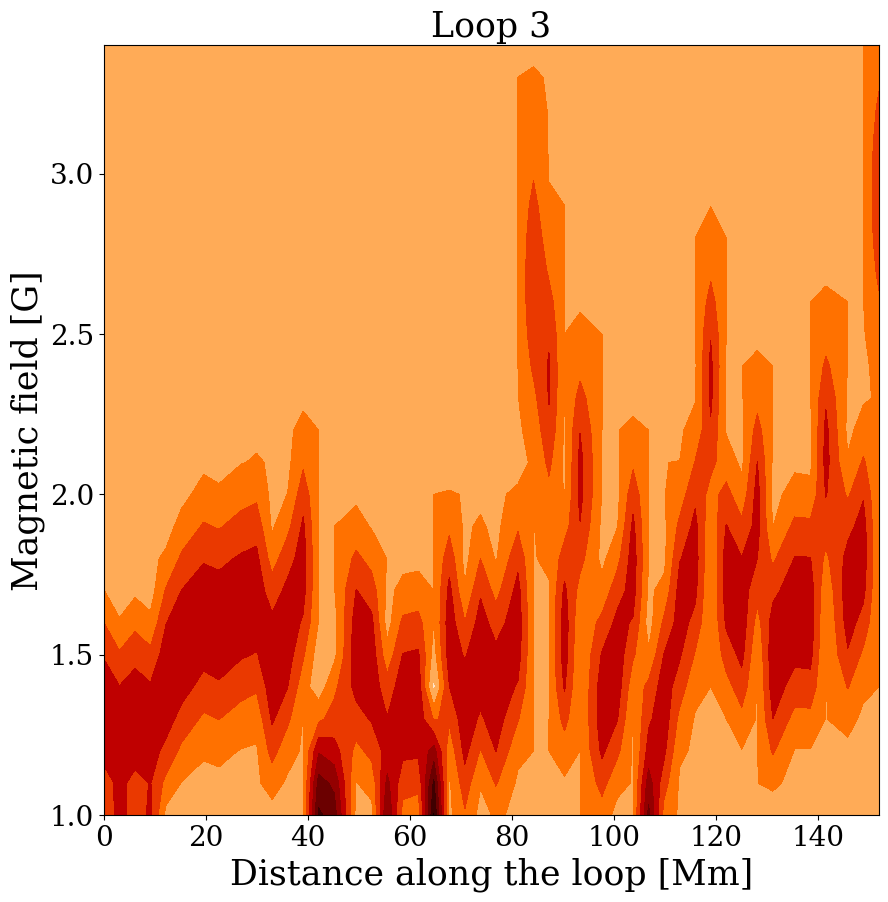}
        \includegraphics[width = 0.23\textwidth]{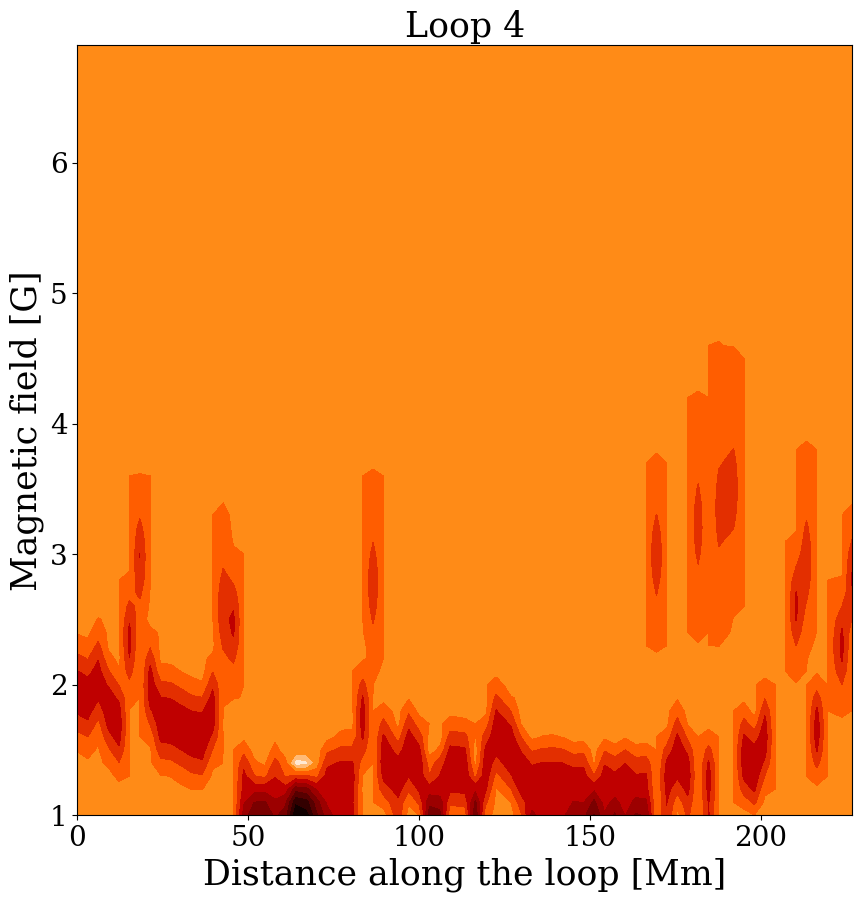}
        \includegraphics[width = 0.23\textwidth]{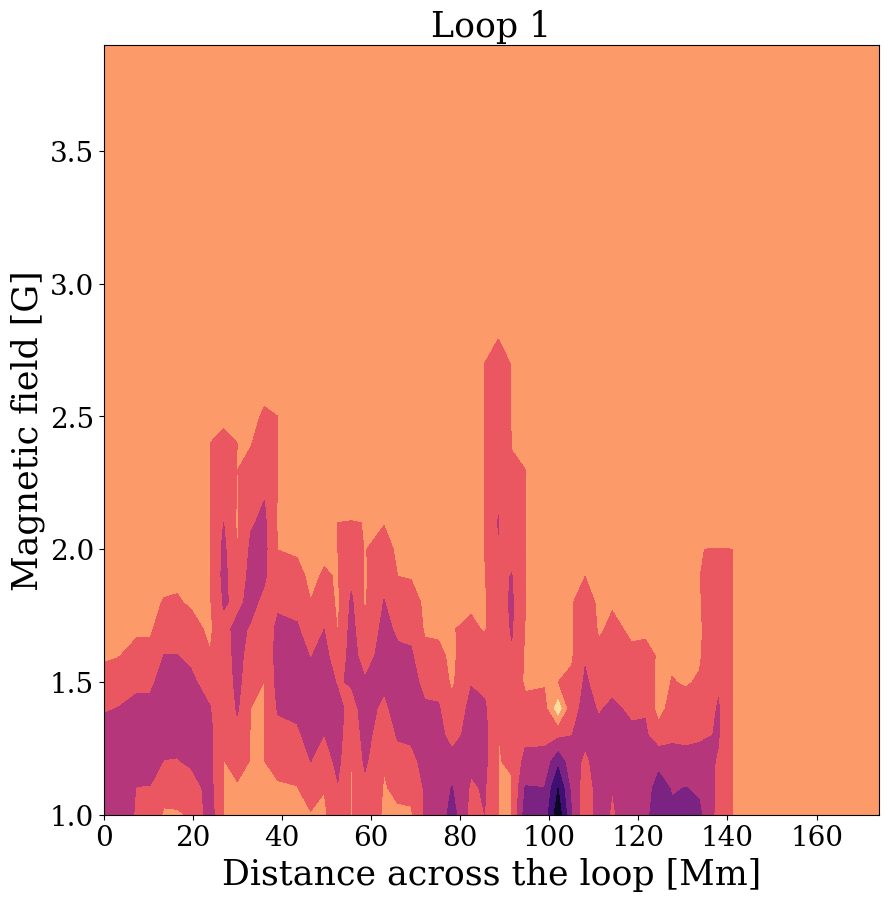}
        \includegraphics[width = 0.23\textwidth]{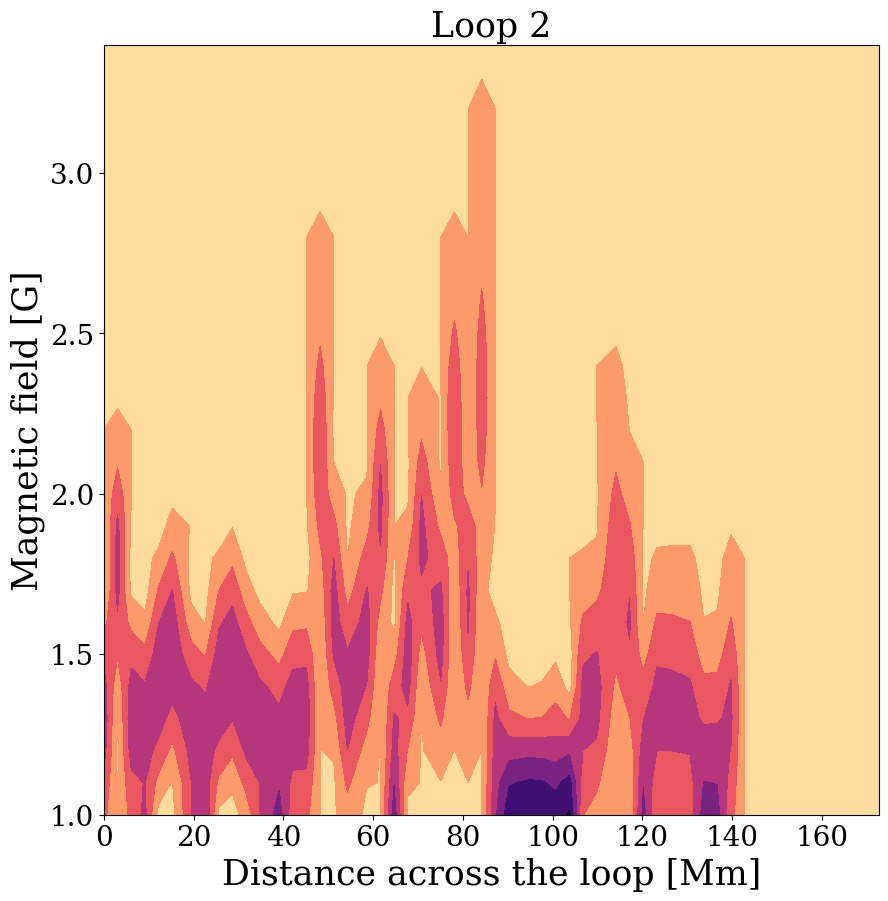}
        \includegraphics[width = 0.23\textwidth]{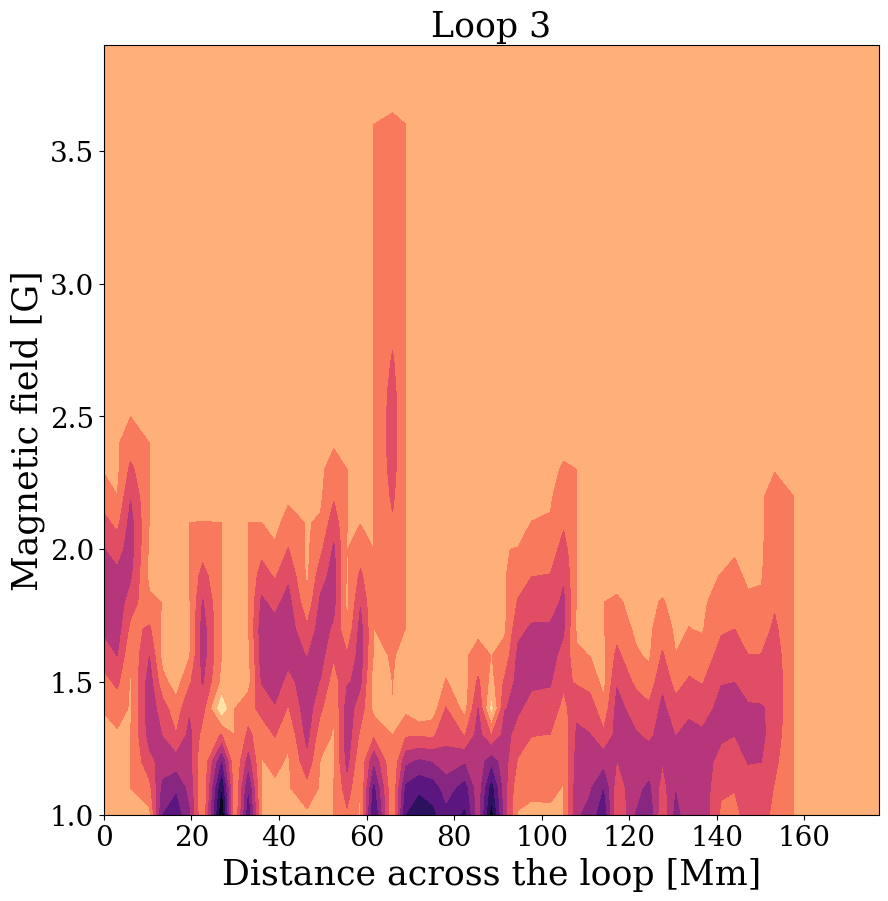}
        \includegraphics[width = 0.23\textwidth]{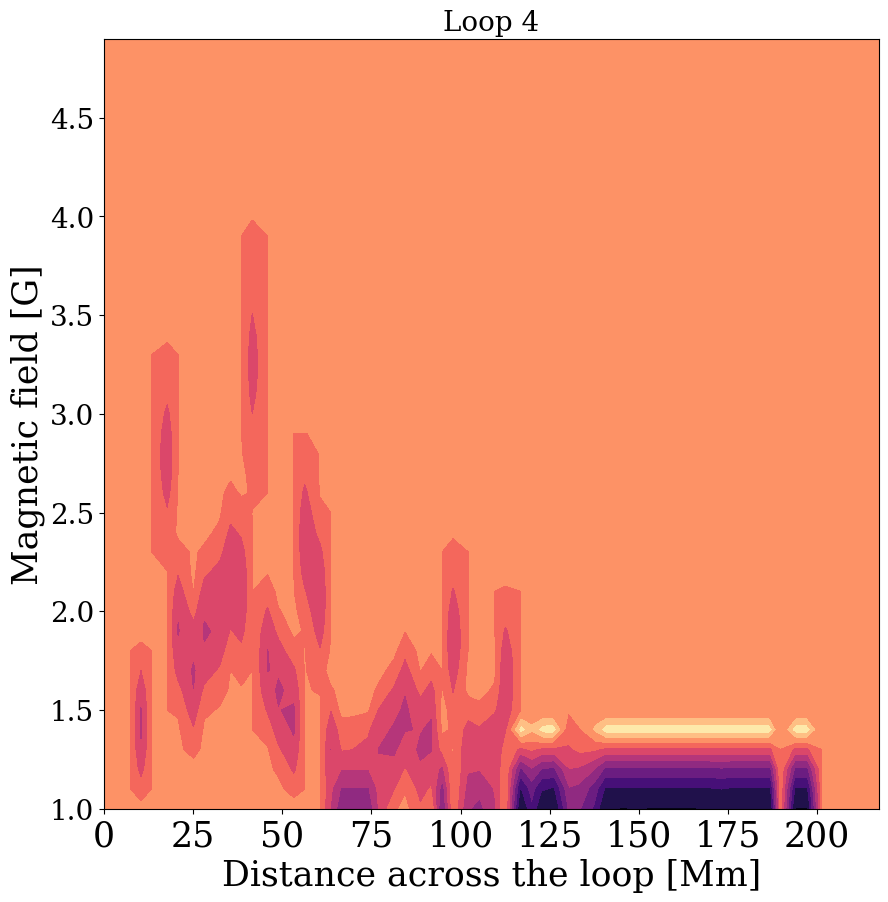}
    \caption{Intensity: Map of the edge enhanced intensity of the FOV at 20:39:09 UT on October 14 2016, along with selected coronal loops and slits across which the variation of the magnetic field is obtained. Variation of coronal magnetic field: In the bottom panels, the first row represents the variation of the coronal magnetic field along the 4 different coronal loops. The bottom panels represent the variation of the coronal magnetic field across these loops.}
    \label{fig:along}
\end{figure*}

Figure \ref{fig:along} displays the full FOV of CoMP in terms of intensity. A spline fitting technique employing 100 spline points is used to obtain smooth arc coordinates along the loops. Different loops at different locations are analyzed, and it was found that the magnetic field is maximum at the footpoints of the coronal loop. This result was also expected from the radial variation of the magnetic field. A Savitzky-Golay filter of $3^{rd}$ polynomial order and window size of 10 pixels is applied along all the loops to smoothen out the variation along the distance further. The middle row of Figure \ref{fig:along} represents the magnetic field variation along the coronal loop. It should also be noted that the magnetic field variation in going from one footpoint to the other in the same loop is not smooth but rather, the magnetic field decreases rapidly and then remains constant for some distance and then increases rapidly. This trend can be appreciated in almost all the selected loops. Here, the colour represents the probability of the magnetic field. An attempt was also made to obtain the magnetic field of coronal loops close to the occulter. But, as evident from the radial variation of the coronal magnetic fields, magnetic fields do not vary much as one moves from footpoints to the loop apex for very small coronal loops near the occulter region. \par
Along the middle of the fitted arc on the coronal loop, a perpendicular slit of 80 pixels is defined. The variation of the magnetic field across this slit is obtained. The distance in the bottom panel of Figure \ref{fig:along} corresponds from inside to outside of the FOV. The magnetic field is expected to increase as it encounters the coronal loop and then decrease as one moves out of that coronal loop and again increases as the slit passes through another coronal loop. A similar trend can also be seen in the bottom panel of Figure \ref{fig:along}.
\section{Summary and Discussion}\label{summary}
A global map of the coronal magnetic field was obtained for the first time by \cite{2020Sci...369..694Y}, and we have applied Bayesian analysis for the first time on the whole global map using the density and associated uncertainty values derived from CoMP observation. We found that the magnetic field obtained using simple inversion does not always match the most probable magnetic field obtained using Bayesian inference. We also showed the impact of choosing two different density priors as well as two different magnetic field priors. The use of Gaussian density priors leads to a better constrained inference of magnetic field strength, because it provides more information to the inference process, in comparison to a uniform density prior over a given range of values. This underscores the crucial significance of obtaining precise density estimates. On the other hand, using a gamma function prior for the magnetic field strength, yields more tightly constrained inferences, although this requires some prior knowledge to set the parameters of the prior probability density. This underscores the crucial significance of obtaining precise density estimates. \par
We have also obtained 3D maps of the global coronal magnetic field, which gives the probability of occurrence of a magnetic field value at each location in the plane of the sky. We have shown the variation of the marginal probability at each location with the varying magnetic field. We also obtained the radial variation of the magnetic field using Bayesian inference and simple inversion. The sharp decrease of the magnetic field with height in AR compared to the QS region is seen in both Bayesian and simple inversion. This result is consistent with \cite{2020ScChE..63.2357Y}. We have also shown the magnetic field variation along and across the coronal loops. We hope this study will be beneficial to understanding the magnetic structure of multi-thermal loops by making use of MHD simulations.\par
Bayesian inference helps us obtain not just a point estimate of the magnetic field but gives a whole distribution of magnetic field values along with their associated probabilities. Our study found that higher magnetic field values are most probable near the base of the coronal loops, and the probability of a high magnetic field decreases while moving outwards in the FOV. This is also expected as the strength of the magnetic field decreases at larger solar radii. \par
Bayesian inference has already been used for seismology of coronal oscillations \citep{2011ApJ...740...44A,2019A&A...625A..35A}, but with assumptions on the density estimation. In this work, we used densities derived from the observations of Fe XIII emission line pairs.\par 
The Bayesian seismology techniques have helped infer parameters in the solar corona and model comparison methods. Bayesian takes care of uncertainty in the parameters' values; this information is passed throughout the analysis from the prior distribution to the posterior and the marginal distributions. One can easily change the prior distribution of density as well as the magnetic field to obtain the more realistic values of the most probable magnetic field. One can also update the value of the magnetic field, density, and phase speed in the prior distribution and likelihood functions, as the information about these parameters keeps on updating with the advancement in observations, as mentioned in \citet{2011ApJ...740...44A}. The reliability of the obtained magnetic field distribution is based on the reliability of the estimated density values. Using Bayesian inference for inversion in upcoming solar facilities such as the Upgraded Coronal Multichannel Polarimeter (UCoMP) and Daniel K. Inouye Solar Telescope (DKIST), we may obtain the magnetic field values with more accuracy.\par
\begin{acknowledgments}
The Mauna Loa Solar Observatory, operated by the High Altitude Observatory as a part of the National Center for Atmospheric Research (NCAR), is acknowledged for providing the data. NCAR is supported by the National Science Foundation. VP is supported by a SERB start-up research grant (File no. SRG/2022/001687). IA is supported by project PID2021-127487NB-I00 from Ministerio de Ciencia, Innovaci\'on y Universidades, and FEDER funds. The python code employed for the MCMC sampling of the posteriors in this study makes use of the {\em emcee} algorithm (Foreman-Mackey et al. 2013) and was developed upon an earlier version created by M. Montes-Sol\'{\i}s. The authors would like to thank Hui Tian and Zihao Yang for sharing the density data. UB and VP express gratitude to ARIES for providing computational resources. Additionally, the insightful suggestions of Dipankar Banerjee and Tom Van Doorsselaere to enhance this work are greatly appreciated. The valuable comments and assistance provided by Ritesh Patel, Bibhuti Kumar Jha, and Nitin Vashishtha are sincerely appreciated. Finally, the authors wish to thank the reviewers for their valuable suggestions.
\end{acknowledgments}

\appendix
\section{INFERENCE WITH A GAMMA PRIOR ON THE MAGNETIC FIELD STRENGTH}\label{AppendixA}
For completeness, we show here results on global maps of the coronal magnetic field, obtained with the use of a Gamma distribution for the prior on magnetic field strength. The distribution is strictly positive and has no upper boundary. A parametric form of this distribution can be expressed as:
\begin{equation}\label{gamma_prior_df}
    p(B) =\begin{cases}
    \frac{\beta^\alpha x^{\alpha-1}e^{-\beta x}}{\Gamma(\alpha)} & \text{$B_{\textit{min}} \le B\le B_{\textit{max}}$} \\ 
    0 & \text{otherwise}.
    \end{cases}
\end{equation}
Here, $\alpha$ and $\beta$ represent the shape and rate parameters. The mean and coefficient of variance of the distribution is given by \(\alpha/\beta\) and \(\alpha/\beta^2\), respectively. To determine the parameters of the distribution, we used information from the results of the simple inversion. In particular, we took the magnetic field obtained from simple inversion as the mean and the magnetic field's uncertainty as the coefficient of variance in the distribution at each location. \par
\begin{figure*}[t]
    \centering
    \includegraphics[height=0.3\textheight]{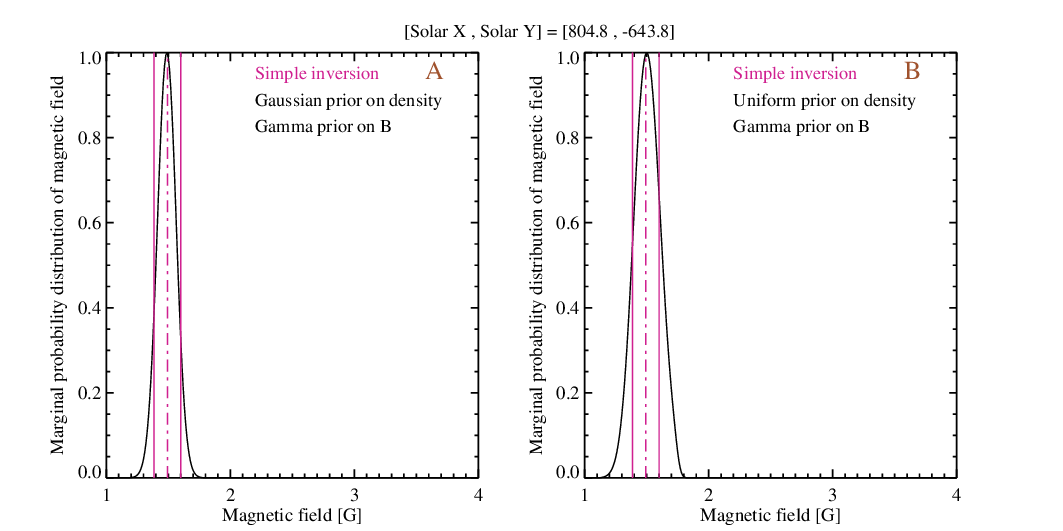}
    \includegraphics[height=0.3\textheight]{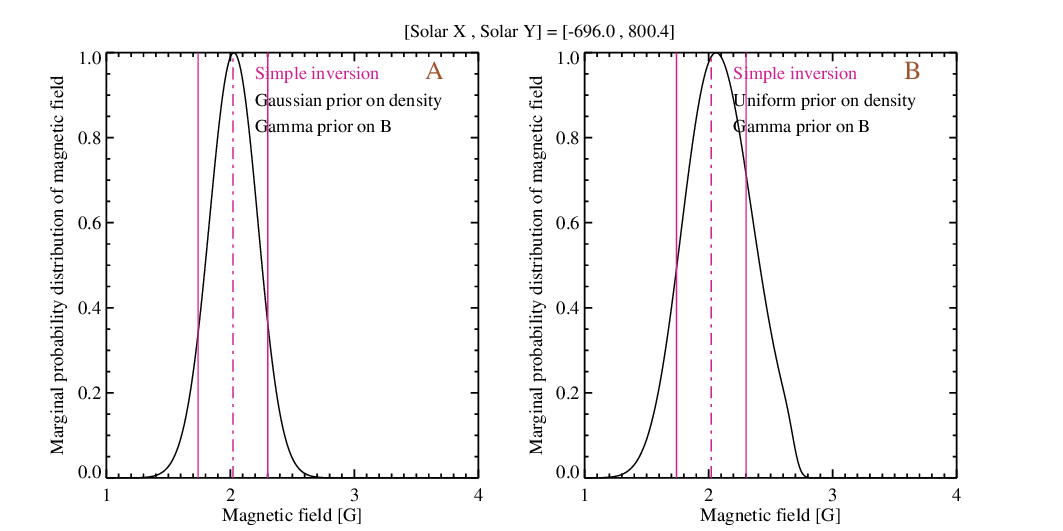}
    \caption{Marginal magnetic field using gamma prior on magnetic field: (A) Marginal magnetic field at two locations using Gaussian prior on density. (B) Marginal magnetic field at the same locations using a uniform prior on density. The vertical dot-dash violet line corresponds to the magnetic field at the same location using simple inversion, and the vertical solid violet line corresponds to the lower and upper limits of the magnetic field. Pixel location and other parameters like phase speed, density, magnetic field and uncertainty associated with these parameters are the same as in Figure \ref{fig:Marginal_at_each_pixel}.}
    \label{fig:Gamma_Marginal_at_each_pixel}
\end{figure*}

We obtained the magnetic field with the gamma prior on the magnetic field and both uniform and Gaussian priors on density. The marginal magnetic field distribution at each pixel for both density priors now exhibits a Gaussian distribution, as depicted in Figure \ref{fig:Gamma_Marginal_at_each_pixel}.\par
\begin{figure*}
    \centering
    \includegraphics[height = 0.4\textheight]{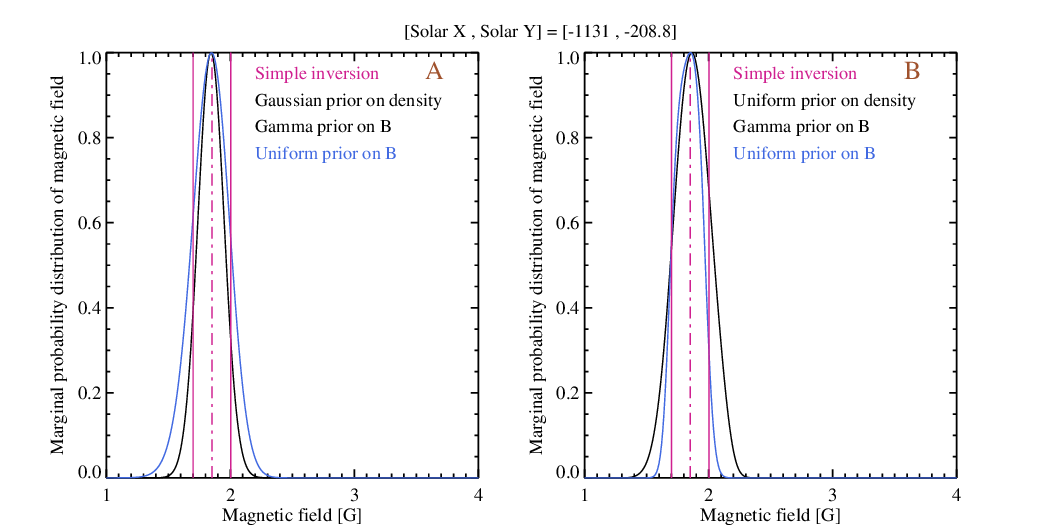}  
    \caption{Comparison of marginal probability distribution: (A) Marginal probability distribution of magnetic field at location (-1131.0,208.8) using Gaussian prior on density. (B) Marginal magnetic field at the same location using a uniform prior on density. The vertical solid and dot-dash lines are the same as in Figure \ref{fig:Marginal_at_each_pixel}. The black and royal blue curves correspond to the gamma prior on the magnetic field and uniform prior on the magnetic field, respectively. Phase speed and associated uncertainty at this location is is \SI{408(12)}{\kilo\meter\per\second} and density and associated uncertainty values are \SI{1.60(0.24)e-13}{\kg\per\cubic\meter}. These results in the magnetic field and associated uncertainty obtained using simple inversion as \SI{1.9(0.2)}{G}.}
    \label{fig:comparison_gamma_uniform_marginal}
\end{figure*}
One comparison between the marginal magnetic field distribution evaluated using a uniform prior on the magnetic field and gamma prior on the magnetic field is also made at location [-1131, -208.8] in Figure \ref{fig:comparison_gamma_uniform_marginal}. The violet lines are as mentioned above. The royal blue curve corresponds to the marginal magnetic field distribution obtained using a uniform prior on the magnetic field, whereas the black curve corresponds to the gamma prior on the magnetic field for both density distributions. It is clear from Figure \ref{fig:comparison_gamma_uniform_marginal} that the marginal magnetic field distribution is more constrained when gamma prior on the magnetic field is used as compared to when uniform prior on the magnetic field is used. \par
\begin{figure}
    \centering
    \includegraphics[height = 6.5cm]{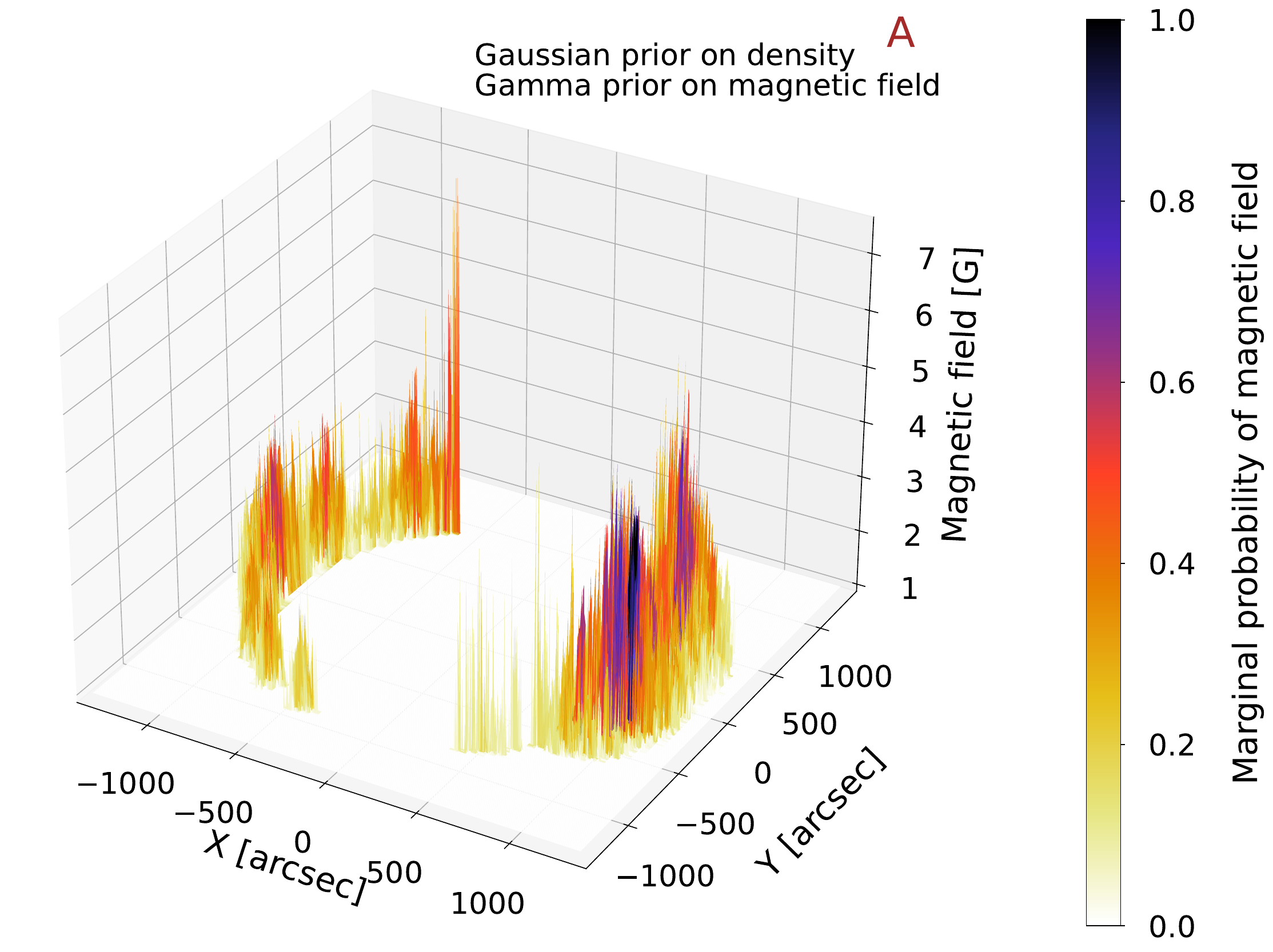}
    \includegraphics[height = 6.5cm]{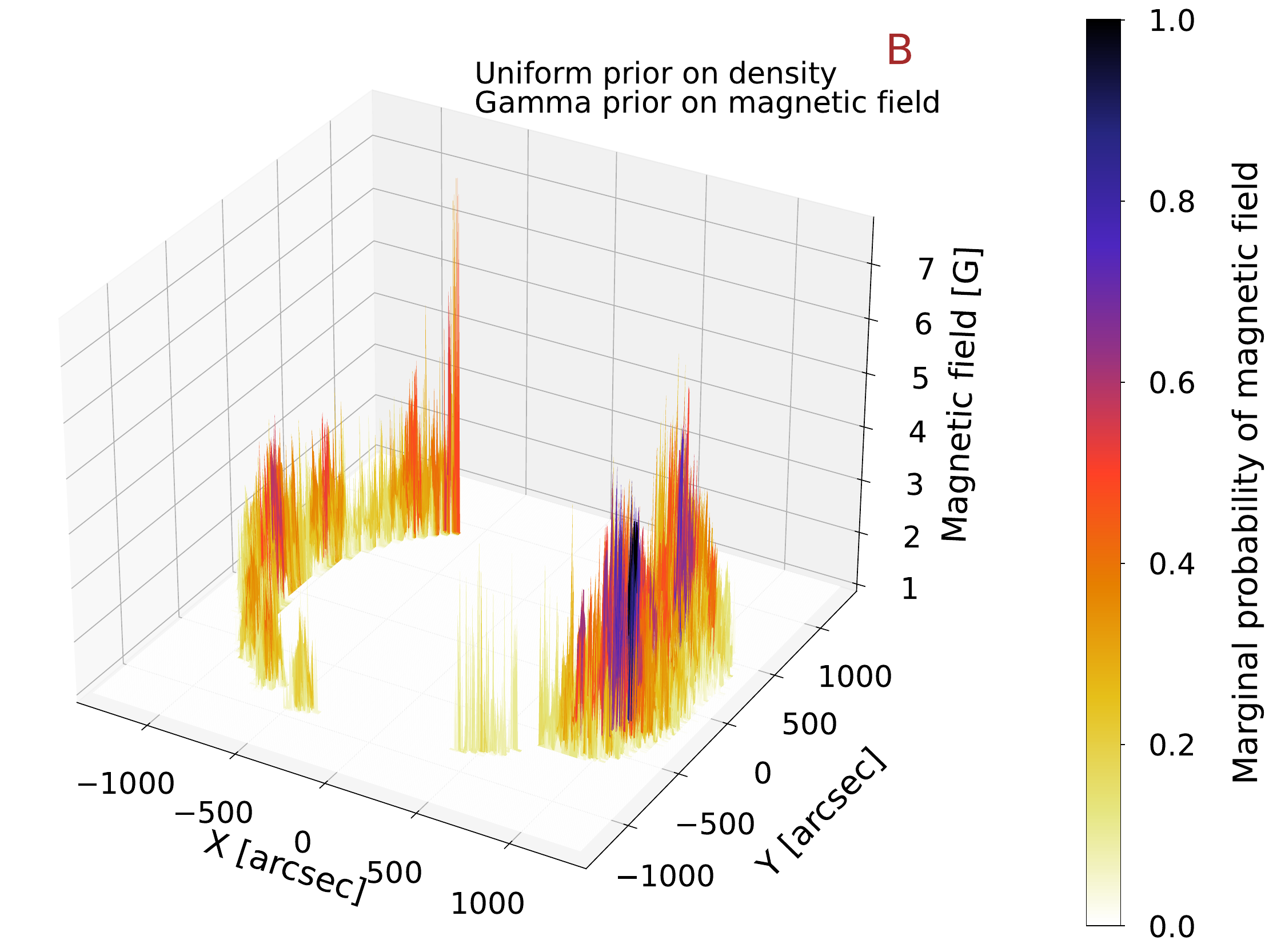}
    \caption{[A] Marginal magnetic field distribution obtained using Gaussian prior density distribution and a gamma prior magnetic field distribution in the whole FOV of CoMP. [B] Marginal distribution of magnetic field obtained using a uniform prior density distribution and a gamma prior magnetic field distribution in the whole FOV of CoMP.}
    \label{fig:marginal_gamma_magnetic_field}
\end{figure}
\begin{figure*}
    \centering
    \includegraphics[height = 0.4\textheight]{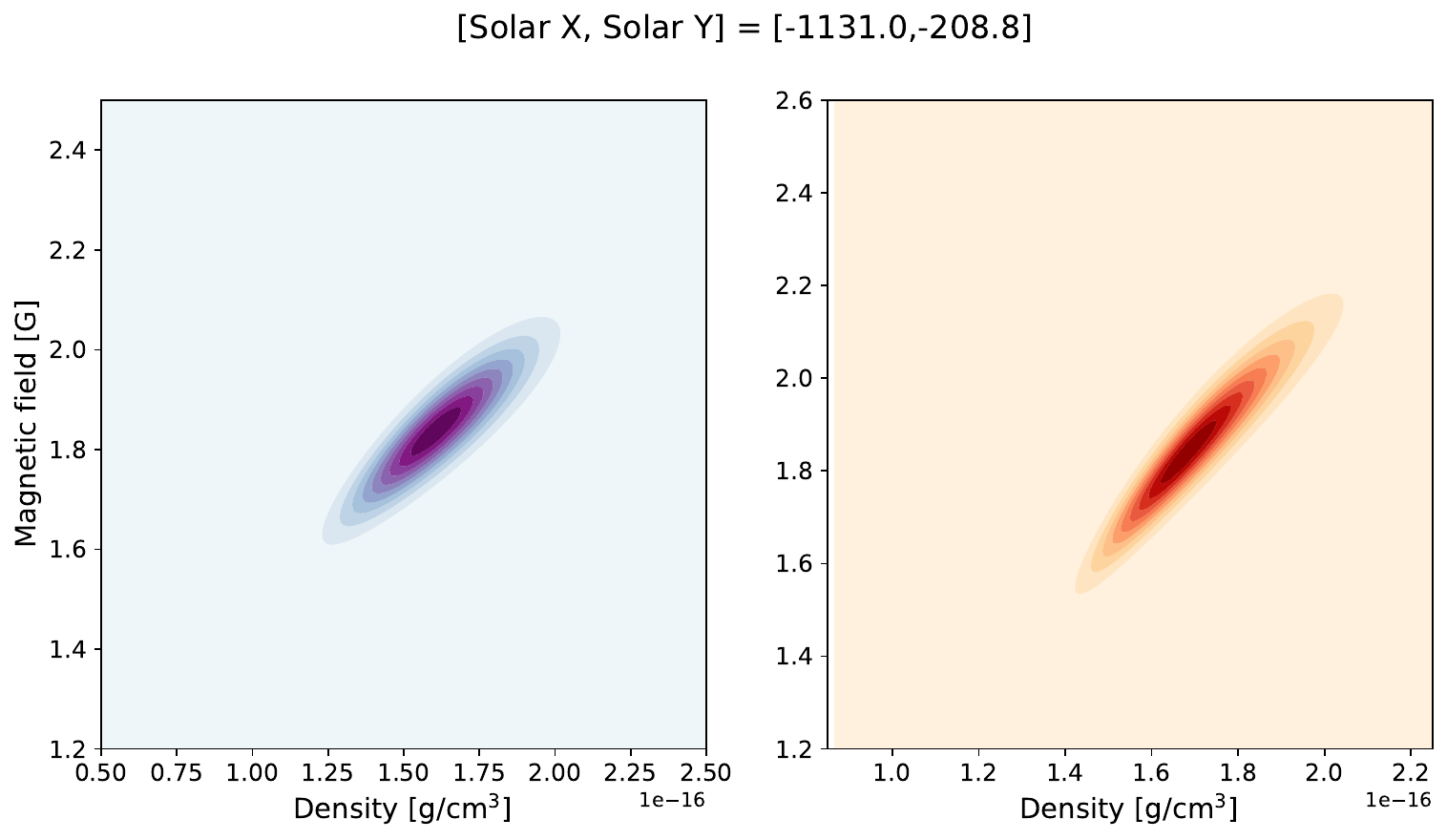}  
    \caption{Joint probability distributions of magnetic field and density at the specified location. For both the priors on density, it is the maximum for which the prior probabilities are also maximum.}
    \label{fig:joint_probability_gamma}
\end{figure*}

The marginal probability distribution of the magnetic field, as obtained through a Gamma prior on the magnetic field, demonstrates more tightly constrained values when compared to the results obtained with a uniform prior on the magnetic field (Figure \ref{fig:Marginal_at_each_pixel}).
Just as in the case of a uniform prior on the magnetic field (Figure \ref{fig:marginal_magnetic_field}), we have also plotted the marginal probability of the magnetic field, using gamma prior on the magnetic field and both the density distribution on density, in the whole field of view. These plots are shown in Figure \ref{fig:marginal_gamma_magnetic_field}. The height and colour of each bar are as explained previously in Section \ref{sec:Bayesian}. The joint probability of magnetic field and density for both the priors on density and gamma prior on magnetic field is shown in Figure \ref{fig:joint_probability_gamma}. Again, the joint probability is the maximum for which the priors on density and magnetic field have a higher probability. \par
\begin{figure*}
    \centering
    \includegraphics[height = 0.4\textheight]{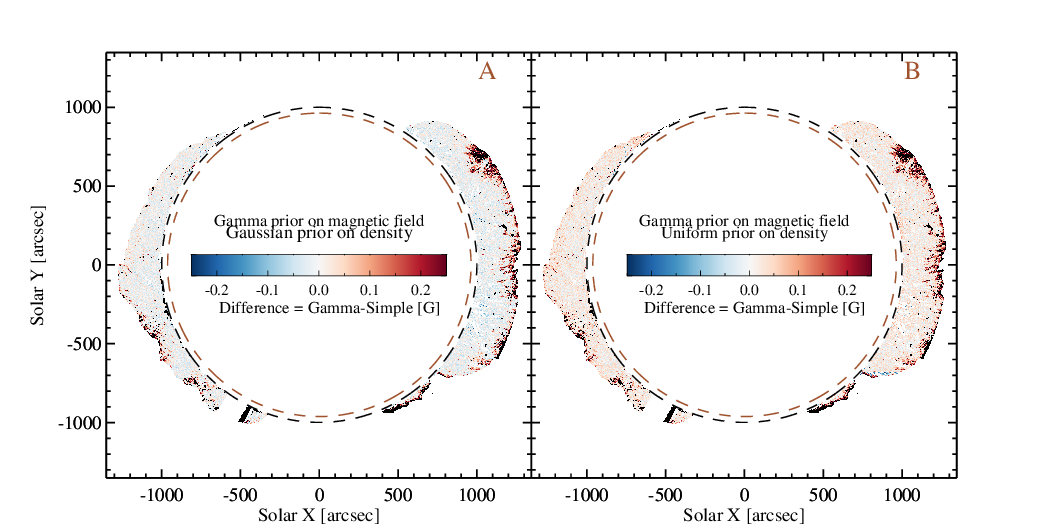}  
    \caption{Magnetic field obtained using gamma prior on magnetic field versus simple inversion: (A) Difference between the most probable value of magnetic field obtained using Bayesian inference and simple inversion at each pixel in CoMP FOV using Gaussian prior on density and gamma prior on magnetic field. (B) Difference between the most probable value of magnetic field obtained using Bayesian inference and simple inversion at the same pixel in CoMP FOV using a uniform prior on density and gamma prior on magnetic field.}
    \label{fig:comparison_gamma_versus_simple}
\end{figure*}

A comparison between the most probable magnetic field obtained using gamma prior on magnetic field for two different density distributions and the magnetic field obtained using simple inversion is also made and depicted in Figure \ref{fig:comparison_gamma_versus_simple}. This also represents that the difference is mainly near the occulter or at the edge of FOV for the abovementioned reasons.

\begin{figure*}
    \centering
    \includegraphics[width = 0.32\textwidth]{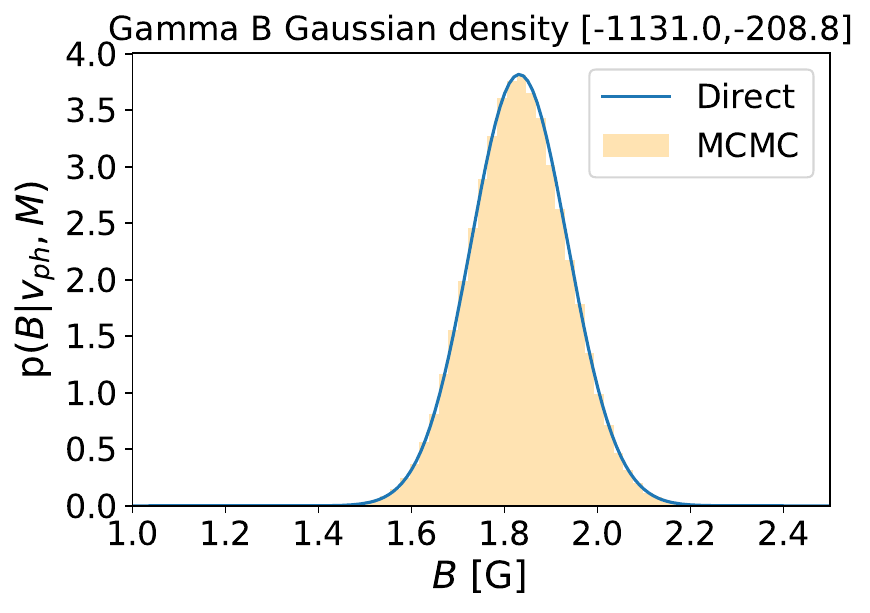}
    \includegraphics[width = 0.32\textwidth]{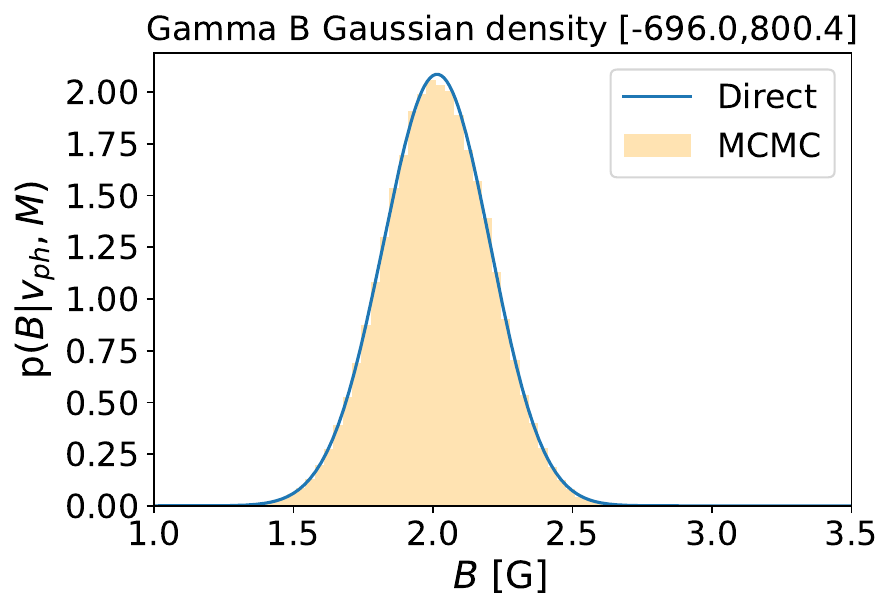}
    \includegraphics[width = 0.32\textwidth]{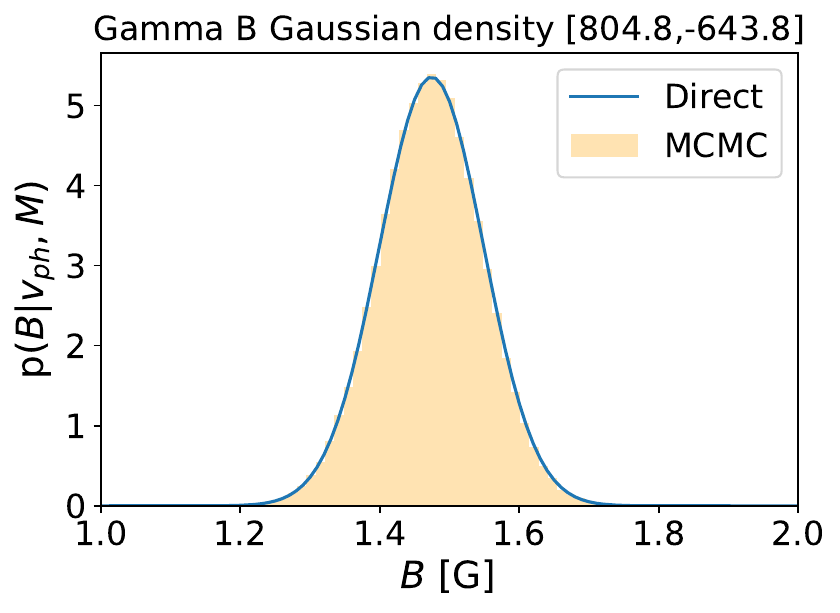}
    \includegraphics[width = 0.32\textwidth]{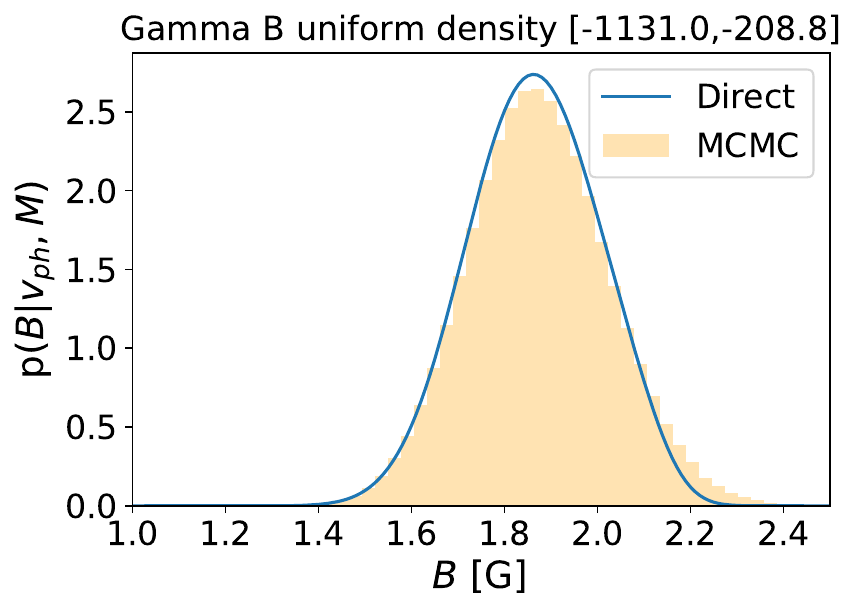}
    \includegraphics[width = 0.32\textwidth]{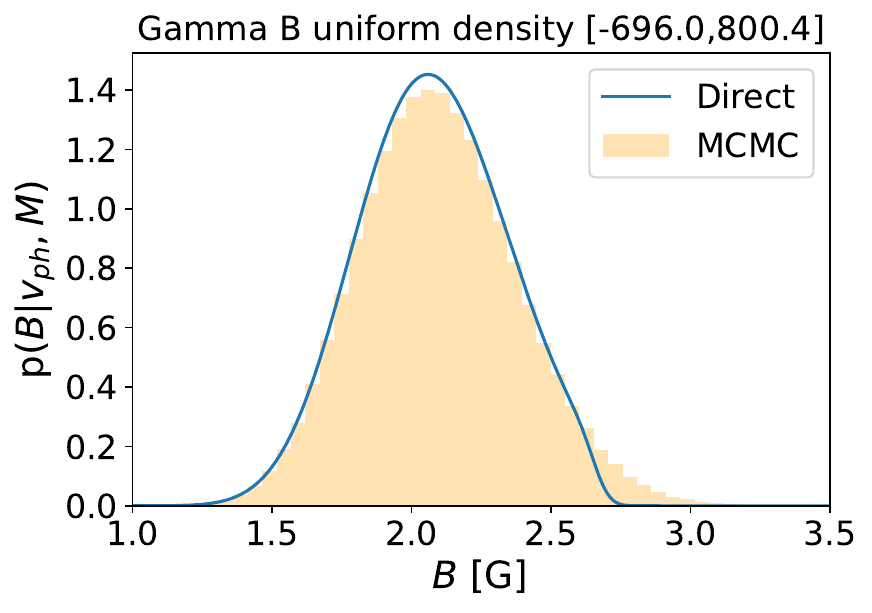}
    \includegraphics[width = 0.32\textwidth]{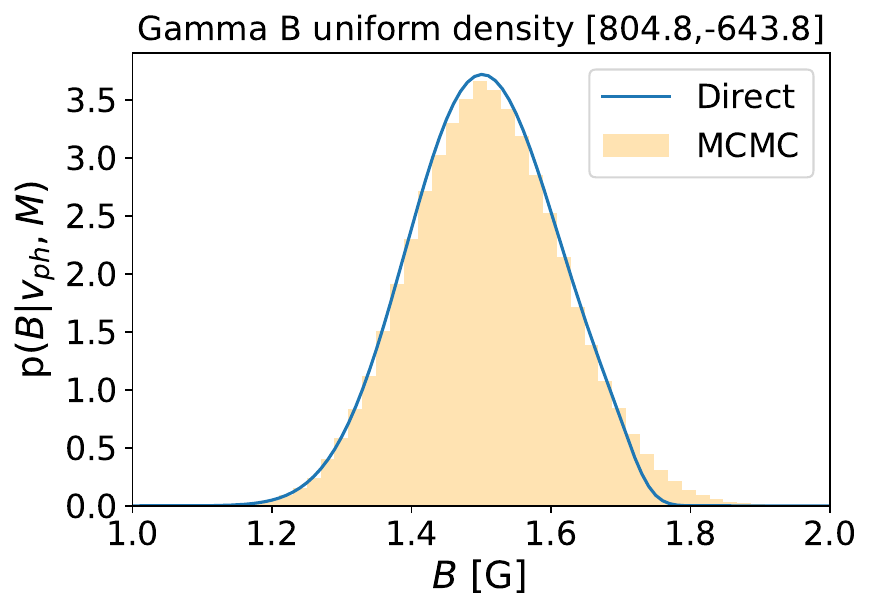}
    \caption{[Top panels] Marginal magnetic field distribution for gamma prior on magnetic field and Gaussian prior on density. [Bottom panels] Marginal magnetic field distribution for gamma prior on magnetic field and uniform prior on density. Orange-coloured Histograms are the results obtained using the \textit{emcee} algorithm of MCMC, whereas the blue solid curve represents the marginal magnetic field distribution obtained using Bayesian inference. The pixel locations are as in Figures \ref{fig:joint_probability} and \ref{fig:Marginal_at_each_pixel}. The marginal probability distributions are not normalized for the sake of comparison with the histograms of samples obtained using MCMC.}
    \label{gamma_magnetic_emcee}
\end{figure*}

\begin{figure*}
    \centering
    \includegraphics[width = 0.32\textwidth]{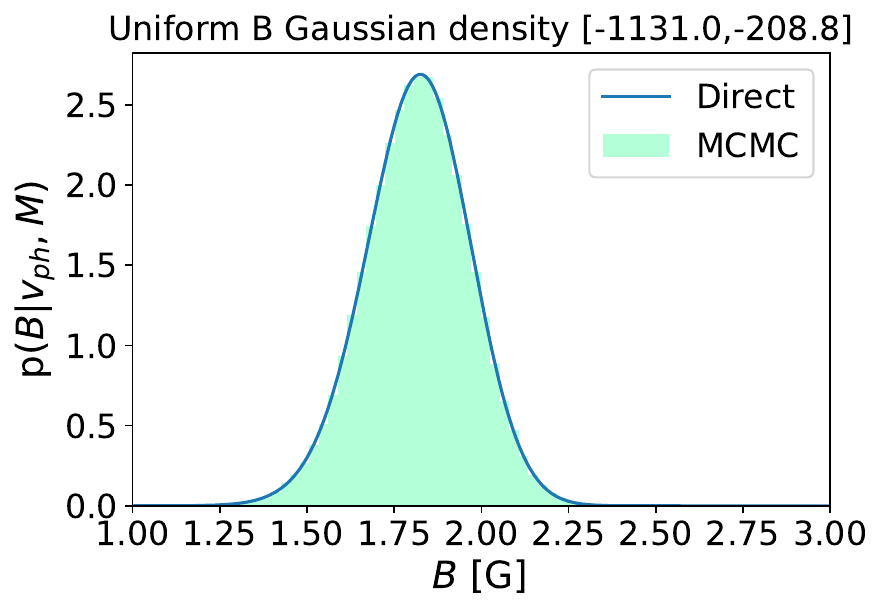}
    \includegraphics[width = 0.32\textwidth]{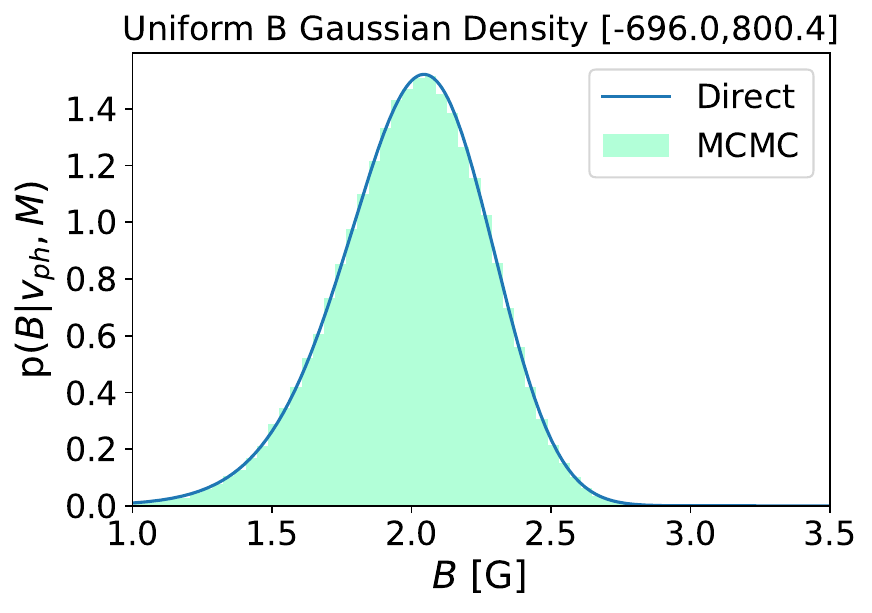}
    \includegraphics[width = 0.32\textwidth]{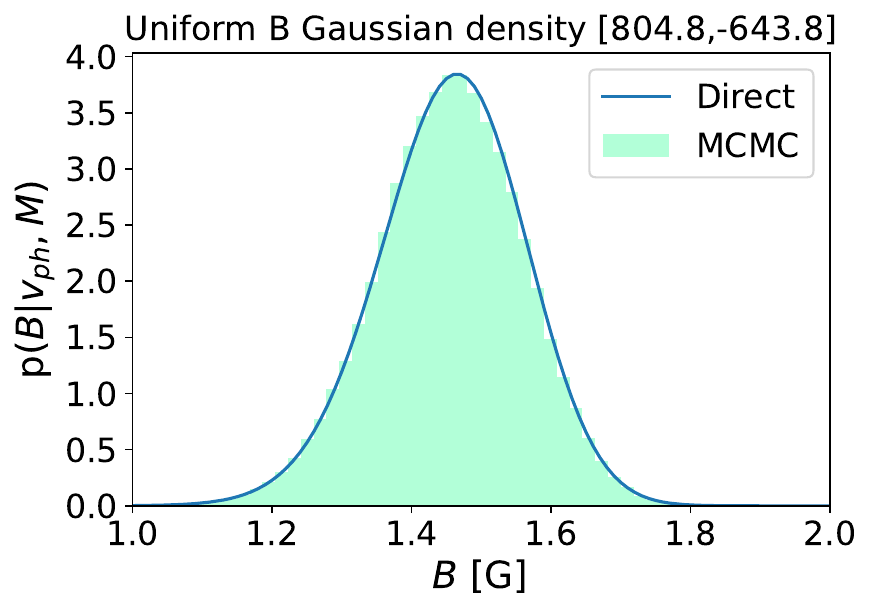}
    \includegraphics[width = 0.32\textwidth]{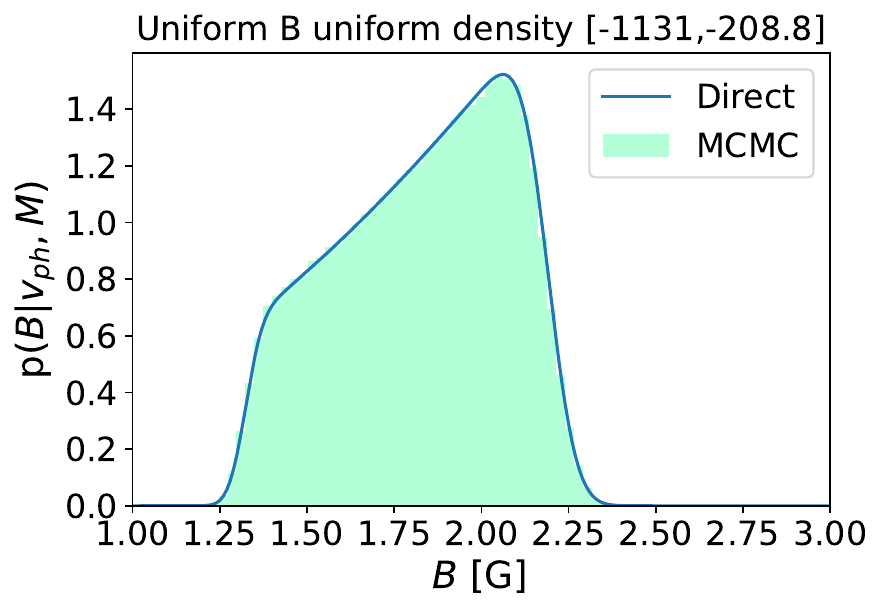}
    \includegraphics[width = 0.32\textwidth]{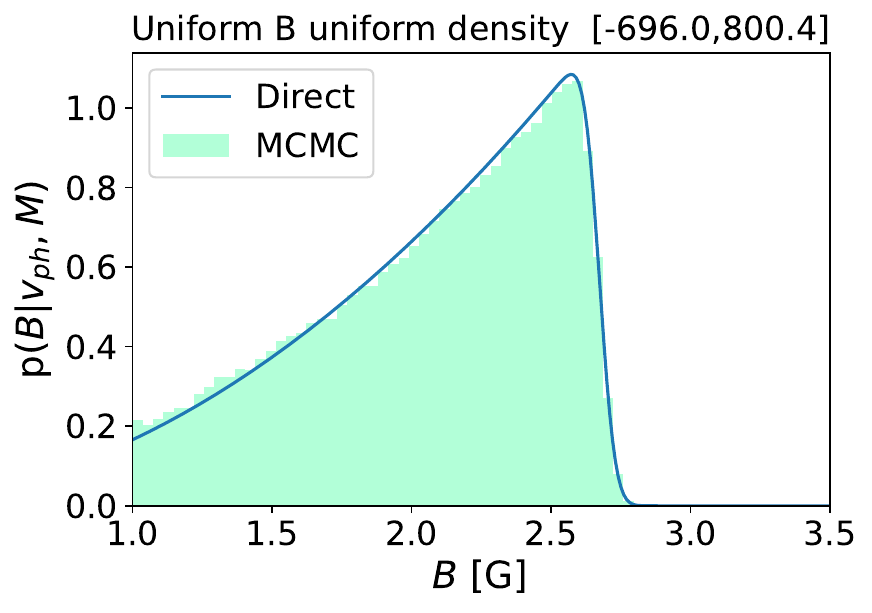}
    \includegraphics[width = 0.32\textwidth]{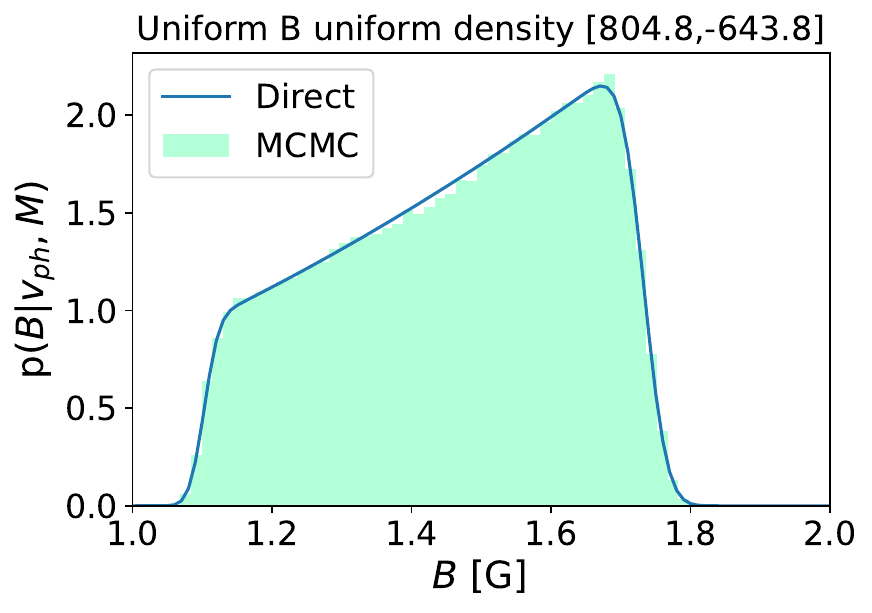}
    \caption{[Top panels] Marginal magnetic field distribution for a uniform prior on magnetic field and Gaussian prior on density. [Bottom panels] Marginal magnetic field distribution for a uniform prior on magnetic field and uniform prior on density. Green-coloured Histograms are the results obtained using the \textit{emcee} algorithm of MCMC, whereas the blue solid curve represents the marginal magnetic field distribution obtained using Bayesian inference. The pixel locations are as in Figures \ref{fig:joint_probability} and \ref{fig:Marginal_at_each_pixel}. The marginal probability distributions are not normalized for the sake of comparison with the histograms of samples obtained using MCMC.}
    \label{uniform_magnetic_emcee}
\end{figure*}

\begin{figure*}[t!]
    \centering
    \includegraphics[height = 0.4\textheight]{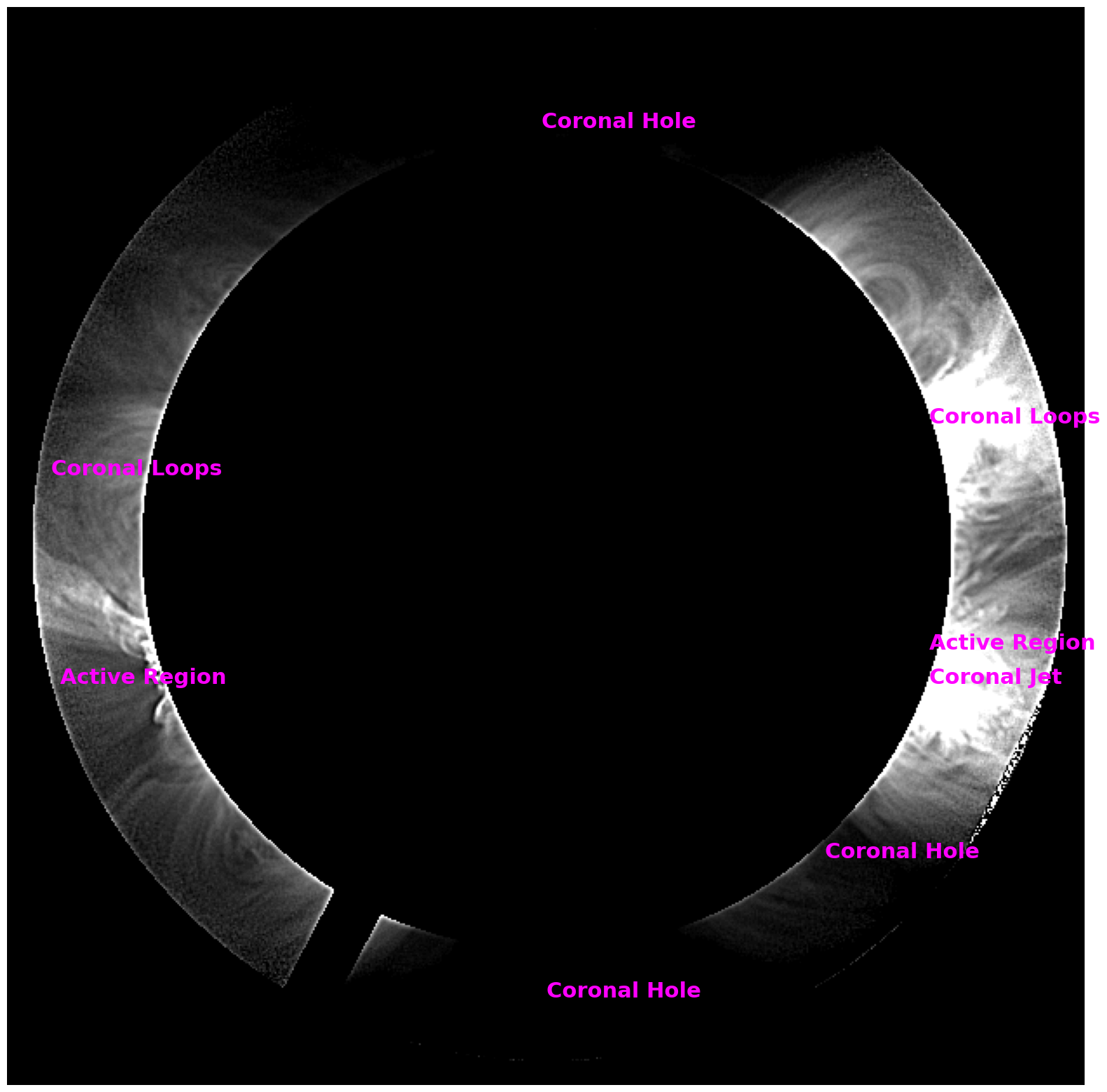}  
    \caption{Notable coronal features in the CoMP FOV during observation.}
    \label{fig:different_features}
\end{figure*}
\section{Application of MCMC on marginal magnetic field}\label{AppendixB}
We also compared the results obtained using direct numerical integration (Bayesian inference) with Markov chain Monte Carlo (MCMC) sampling for the posterior, making use of the \textit{emcee} algorithm \citep{2013PASP..125..306F}. The application of the \textit{emcee} algorithm is explained in \cite{2017ApJ...846...89M} and \cite{2019A&A...625A..35A}. Since there are only two parameters involved as $theta = [B, \rho]$, thus, the number of dimensions in emcee is 2. We used around 15 walkers with 50,000 steps in each walker for \textit{emcee} algorithm. Out of which, we have used an initial 20$\%$ of the iterations for the burn-in phase to avoid any outliers. The marginal posterior distribution of the magnetic field results in the same distribution for two different density distributions when the gamma distribution is taken as the prior distribution on the magnetic field. Further, one can check from Figure \ref{gamma_magnetic_emcee} that there is no effect of change in the prior distribution of density on the marginal probability distribution of the magnetic field as mentioned in \cite{2019A&A...625A..35A}. Also, the marginal probability distribution of the magnetic field is the same even if we consider two different priors of the magnetic field (gamma and uniform), keeping the density distribution the same (Gaussian prior on density). This can be seen from the top panels of Figures \ref{gamma_magnetic_emcee} and \ref{uniform_magnetic_emcee}. We have compared the results with MCMC only for the locations shown in Figures \ref{fig:joint_probability} and \ref{fig:Marginal_at_each_pixel}. The results obtained using MCMC and direct numerical integration also match the case when uniform prior is considered for density and the magnetic field. 

\section{Features in FOV}\label{AppendixC}
Various structures, including coronal holes, active regions, and more, can be observed in the CoMP intensity images. Coronal holes are notably present in polar regions and along the west limb. The locations of these various coronal features are shown in Figure \ref{fig:different_features}.
\bibliography{sample631}{}

\begin{thebibliography}{}
\expandafter\ifx\csname natexlab\endcsname\relax\def\natexlab#1{#1}\fi
\providecommand{\url}[1]{\href{#1}{#1}}
\providecommand{\dodoi}[1]{doi:~\href{http://doi.org/#1}{\nolinkurl{#1}}}
\providecommand{\doeprint}[1]{\href{http://ascl.net/#1}{\nolinkurl{http://ascl.net/#1}}}
\providecommand{\doarXiv}[1]{\href{https://arxiv.org/abs/#1}{\nolinkurl{https://arxiv.org/abs/#1}}}

\bibitem[{{Altschuler} \& {Newkirk}(1969)}]{1969SoPh....9..131A}
{Altschuler}, M.~D., \& {Newkirk}, G. 1969, \solphys, 9, 131,
  \dodoi{10.1007/BF00145734}

\bibitem[{{Arregui}(2012)}]{2012ASSP...33..159A}
{Arregui}, I. 2012, in Astrophysics and Space Science Proceedings, Vol.~33,
  Multi-scale Dynamical Processes in Space and Astrophysical Plasmas, 159,
  \dodoi{10.1007/978-3-642-30442-2_18}

\bibitem[{{Arregui}(2015)}]{2015RSPTA.37340261A}
{Arregui}, I. 2015, Philosophical Transactions of the Royal Society of London
  Series A, 373, 20140261, \dodoi{10.1098/rsta.2014.0261}

\bibitem[{{Arregui}(2018)}]{2018AdSpR..61..655A}
---. 2018, Advances in Space Research, 61, 655,
  \dodoi{10.1016/j.asr.2017.09.031}

\bibitem[{{Arregui}(2021)}]{2021ApJ...915L..25A}
---. 2021, \apjl, 915, L25, \dodoi{10.3847/2041-8213/ac0d53}

\bibitem[{Arregui(2022)}]{10.3389/fspas.2022.826947}
Arregui, I. 2022, Frontiers in Astronomy and Space Sciences, 9,
  \dodoi{10.3389/fspas.2022.826947}

\bibitem[{{Arregui} {et~al.}(2012){Arregui}, {Ballester}, {Goossens}, {Oliver},
  \& {Ramos}}]{2012ASPC..456..121A}
{Arregui}, I.~., {Ballester}, J., {Goossens}, M., {Oliver}, R., \& {Ramos}, A.
  2012, in Astronomical Society of the Pacific Conference Series, Vol. 456,
  Fifth Hinode Science Meeting, ed. L.~{Golub}, I.~{De Moortel}, \&
  T.~{Shimizu}, 121

\bibitem[{{Arregui} \& {Asensio Ramos}(2011)}]{2011ApJ...740...44A}
{Arregui}, I., \& {Asensio Ramos}, A. 2011, \apj, 740, 44,
  \dodoi{10.1088/0004-637X/740/1/44}

\bibitem[{{Arregui} \& {Goossens}(2019)}]{2019A&A...622A..44A}
{Arregui}, I., \& {Goossens}, M. 2019, \aap, 622, A44,
  \dodoi{10.1051/0004-6361/201833813}

\bibitem[{{Arregui} {et~al.}(2019){Arregui}, {Montes-Sol{\'\i}s}, \& {Asensio
  Ramos}}]{2019A&A...625A..35A}
{Arregui}, I., {Montes-Sol{\'\i}s}, M., \& {Asensio Ramos}, A. 2019, \aap, 625,
  A35, \dodoi{10.1051/0004-6361/201834324}

\bibitem[{Arregui {et~al.}(2018)Arregui, Oliver, \& Ballester}]{article}
Arregui, I., Oliver, R., \& Ballester, J. 2018, Living Reviews in Solar
  Physics, 15, \dodoi{10.1007/s41116-018-0012-6}

\bibitem[{Arregui \& Ramos(2011)}]{Arregui_2011}
Arregui, I., \& Ramos, A.~A. 2011, The Astrophysical Journal, 740, 44,
  \dodoi{10.1088/0004-637x/740/1/44}

\bibitem[{{Arregui} {et~al.}(2014){Arregui}, {Ramos}, \&
  {D{\'\i}az}}]{2014IAUS..300..393A}
{Arregui}, I., {Ramos}, A.~A., \& {D{\'\i}az}, A.~J. 2014, in Nature of
  Prominences and their Role in Space Weather, ed. B.~{Schmieder}, J.-M.
  {Malherbe}, \& S.~T. {Wu}, Vol. 300, 393--394,
  \dodoi{10.1017/S1743921313011241}

\bibitem[{Aschwanden(2006)}]{aschwanden2006coronal}
Aschwanden, M.~J. 2006, Philosophical Transactions of the Royal Society A:
  Mathematical, Physical and Engineering Sciences, 364, 417

\bibitem[{{Aschwanden} {et~al.}(2002){Aschwanden}, {de Pontieu}, {Schrijver},
  \& {Title}}]{2002SoPh..206...99A}
{Aschwanden}, M.~J., {de Pontieu}, B., {Schrijver}, C.~J., \& {Title}, A.~M.
  2002, \solphys, 206, 99, \dodoi{10.1023/A:1014916701283}

\bibitem[{{Aschwanden} {et~al.}(1999){Aschwanden}, {Fletcher}, {Schrijver}, \&
  {Alexander}}]{1999ApJ...520..880A}
{Aschwanden}, M.~J., {Fletcher}, L., {Schrijver}, C.~J., \& {Alexander}, D.
  1999, \apj, 520, 880, \dodoi{10.1086/307502}

\bibitem[{{Aschwanden} \& {Schrijver}(2011)}]{2011ApJ...736..102A}
{Aschwanden}, M.~J., \& {Schrijver}, C.~J. 2011, \apj, 736, 102,
  \dodoi{10.1088/0004-637X/736/2/102}

\bibitem[{{Banerjee} {et~al.}(2007){Banerjee}, {Erd{\'e}lyi}, {Oliver}, \&
  {O'Shea}}]{2007SoPh..246....3B}
{Banerjee}, D., {Erd{\'e}lyi}, R., {Oliver}, R., \& {O'Shea}, E. 2007,
  \solphys, 246, 3, \dodoi{10.1007/s11207-007-9029-z}

\bibitem[{{Bayes} \& {Price}(1763)}]{1763RSPT...53..370B}
{Bayes}, M., \& {Price}, M. 1763, Philosophical Transactions of the Royal
  Society of London Series I, 53, 370, \dodoi{10.1098/rstl.1763.0053}

\bibitem[{{De Moortel} \& {Nakariakov}(2012)}]{2012RSPTA.370.3193D}
{De Moortel}, I., \& {Nakariakov}, V.~M. 2012, Philosophical Transactions of
  the Royal Society of London Series A, 370, 3193,
  \dodoi{10.1098/rsta.2011.0640}

\bibitem[{{De Pontieu} {et~al.}(2007){De Pontieu}, {McIntosh}, {Carlsson},
  {Hansteen}, {Tarbell}, {Schrijver}, {Title}, {Shine}, {Tsuneta}, {Katsukawa},
  {Ichimoto}, {Suematsu}, {Shimizu}, \& {Nagata}}]{2007Sci...318.1574D}
{De Pontieu}, B., {McIntosh}, S.~W., {Carlsson}, M., {et~al.} 2007, Science,
  318, 1574, \dodoi{10.1126/science.1151747}

\bibitem[{{Dere} {et~al.}(2019){Dere}, {Del Zanna}, {Young}, {Landi}, \&
  {Sutherland}}]{2019ApJS..241...22D}
{Dere}, K.~P., {Del Zanna}, G., {Young}, P.~R., {Landi}, E., \& {Sutherland},
  R.~S. 2019, \apjs, 241, 22, \dodoi{10.3847/1538-4365/ab05cf}

\bibitem[{{Dere} {et~al.}(1997){Dere}, {Landi}, {Mason}, {Monsignori Fossi}, \&
  {Young}}]{1997A&AS..125..149D}
{Dere}, K.~P., {Landi}, E., {Mason}, H.~E., {Monsignori Fossi}, B.~C., \&
  {Young}, P.~R. 1997, \aaps, 125, 149, \dodoi{10.1051/aas:1997368}

\bibitem[{{Domingo} {et~al.}(1995){Domingo}, {Fleck}, \&
  {Poland}}]{1995SoPh..162....1D}
{Domingo}, V., {Fleck}, B., \& {Poland}, A.~I. 1995, \solphys, 162, 1,
  \dodoi{10.1007/BF00733425}

\bibitem[{{Foreman-Mackey} {et~al.}(2013){Foreman-Mackey}, {Hogg}, {Lang}, \&
  {Goodman}}]{2013PASP..125..306F}
{Foreman-Mackey}, D., {Hogg}, D.~W., {Lang}, D., \& {Goodman}, J. 2013, \pasp,
  125, 306, \dodoi{10.1086/670067}

\bibitem[{{Goossens} {et~al.}(2009){Goossens}, {Terradas}, {Andries},
  {Arregui}, \& {Ballester}}]{2009A&A...503..213G}
{Goossens}, M., {Terradas}, J., {Andries}, J., {Arregui}, I., \& {Ballester},
  J.~L. 2009, \aap, 503, 213, \dodoi{10.1051/0004-6361/200912399}

\bibitem[{{Handy} {et~al.}(1999){Handy}, {Acton}, {Kankelborg}, {Wolfson},
  {Akin}, {Bruner}, {Caravalho}, {Catura}, {Chevalier}, {Duncan}, {Edwards},
  {Feinstein}, {Freeland}, {Friedlaender}, {Hoffmann}, {Hurlburt}, {Jurcevich},
  {Katz}, {Kelly}, {Lemen}, {Levay}, {Lindgren}, {Mathur}, {Meyer}, {Morrison},
  {Morrison}, {Nightingale}, {Pope}, {Rehse}, {Schrijver}, {Shine}, {Shing},
  {Strong}, {Tarbell}, {Title}, {Torgerson}, {Golub}, {Bookbinder}, {Caldwell},
  {Cheimets}, {Davis}, {Deluca}, {McMullen}, {Warren}, {Amato}, {Fisher},
  {Maldonado}, \& {Parkinson}}]{1999SoPh..187..229H}
{Handy}, B.~N., {Acton}, L.~W., {Kankelborg}, C.~C., {et~al.} 1999, \solphys,
  187, 229, \dodoi{10.1023/A:1005166902804}

\bibitem[{{He} {et~al.}(2011){He}, {Wang}, \& {Yan}}]{2011JGRA..116.1101H}
{He}, H., {Wang}, H., \& {Yan}, Y. 2011, Journal of Geophysical Research (Space
  Physics), 116, A01101, \dodoi{10.1029/2010JA015610}

\bibitem[{Huang {et~al.}(2018)Huang, Xia, Nelson, Liu, Wiegelmann, Tian,
  Klimchuk, Chen, \& Li}]{Huang_2018}
Huang, Z., Xia, L., Nelson, C.~J., {et~al.} 2018, The Astrophysical Journal,
  854, 80, \dodoi{10.3847/1538-4357/aaa9ba}

\bibitem[{Jess {et~al.}(2016)Jess, Reznikova, Ryans, Christian, Keys,
  Mathioudakis, Mackay, Krishna~Prasad, Banerjee, Grant,
  {et~al.}}]{jess2016solar}
Jess, D.~B., Reznikova, V.~E., Ryans, R.~S., {et~al.} 2016, Nature Physics, 12,
  179

\bibitem[{{Jess} {et~al.}(2016){Jess}, {Reznikova}, {Ryans}, {Christian},
  {Keys}, {Mathioudakis}, {Mackay}, {Krishna Prasad}, {Banerjee}, {Grant},
  {Yau}, \& {Diamond}}]{2016NatPh..12..179J}
{Jess}, D.~B., {Reznikova}, V.~E., {Ryans}, R. S.~I., {et~al.} 2016, Nature
  Physics, 12, 179, \dodoi{10.1038/nphys3544}

\bibitem[{{Kosugi} {et~al.}(2007){Kosugi}, {Matsuzaki}, {Sakao}, {Shimizu},
  {Sone}, {Tachikawa}, {Hashimoto}, {Minesugi}, {Ohnishi}, {Yamada}, {Tsuneta},
  {Hara}, {Ichimoto}, {Suematsu}, {Shimojo}, {Watanabe}, {Shimada}, {Davis},
  {Hill}, {Owens}, {Title}, {Culhane}, {Harra}, {Doschek}, \&
  {Golub}}]{2007SoPh..243....3K}
{Kosugi}, T., {Matsuzaki}, K., {Sakao}, T., {et~al.} 2007, \solphys, 243, 3,
  \dodoi{10.1007/s11207-007-9014-6}

\bibitem[{{Kumari} {et~al.}(2019){Kumari}, {Ramesh}, {Kathiravan}, {Wang}, \&
  {Gopalswamy}}]{2019ApJ...881...24K}
{Kumari}, A., {Ramesh}, R., {Kathiravan}, C., {Wang}, T.~J., \& {Gopalswamy},
  N. 2019, \apj, 881, 24, \dodoi{10.3847/1538-4357/ab2adf}

\bibitem[{Lee {et~al.}(1999)Lee, White, Kundu, Miki{\'{c}}, \&
  McClymont}]{Lee_1999}
Lee, J., White, S.~M., Kundu, M.~R., Miki{\'{c}}, Z., \& McClymont, A.~N. 1999,
  The Astrophysical Journal, 510, 413, \dodoi{10.1086/306556}

\bibitem[{{Long} {et~al.}(2017){Long}, {Valori}, {P{\'e}rez-Su{\'a}rez},
  {Morton}, \& {V{\'a}squez}}]{2017A&A...603A.101L}
{Long}, D.~M., {Valori}, G., {P{\'e}rez-Su{\'a}rez}, D., {Morton}, R.~J., \&
  {V{\'a}squez}, A.~M. 2017, \aap, 603, A101,
  \dodoi{10.1051/0004-6361/201730413}

\bibitem[{{Montes-Sol{\'\i}s} \& {Arregui}(2017)}]{2017ApJ...846...89M}
{Montes-Sol{\'\i}s}, M., \& {Arregui}, I. 2017, \apj, 846, 89,
  \dodoi{10.3847/1538-4357/aa84b7}

\bibitem[{{Morton} {et~al.}(2015){Morton}, {Tomczyk}, \&
  {Pinto}}]{2015NatCo...6.7813M}
{Morton}, R.~J., {Tomczyk}, S., \& {Pinto}, R. 2015, Nature Communications, 6,
  7813, \dodoi{10.1038/ncomms8813}

\bibitem[{{Morton} {et~al.}(2016){Morton}, {Tomczyk}, \&
  {Pinto}}]{2016ApJ...828...89M}
{Morton}, R.~J., {Tomczyk}, S., \& {Pinto}, R.~F. 2016, \apj, 828, 89,
  \dodoi{10.3847/0004-637X/828/2/89}

\bibitem[{{Morton} {et~al.}(2019){Morton}, {Weberg}, \&
  {McLaughlin}}]{2019NatAs...3..223M}
{Morton}, R.~J., {Weberg}, M.~J., \& {McLaughlin}, J.~A. 2019, Nature
  Astronomy, 3, 223, \dodoi{10.1038/s41550-018-0668-9}

\bibitem[{Nakariakov(2000)}]{doi:10.1063/1.1324949}
Nakariakov, V.~M. 2000, AIP Conference Proceedings, 537, 264,
  \dodoi{10.1063/1.1324949}

\bibitem[{{Nakariakov} \& {Ofman}(2001)}]{2001A&A...372L..53N}
{Nakariakov}, V.~M., \& {Ofman}, L. 2001, \aap, 372, L53,
  \dodoi{10.1051/0004-6361:20010607}

\bibitem[{{Nakariakov} {et~al.}(1999){Nakariakov}, {Ofman}, {Deluca},
  {Roberts}, \& {Davila}}]{1999Sci...285..862N}
{Nakariakov}, V.~M., {Ofman}, L., {Deluca}, E.~E., {Roberts}, B., \& {Davila},
  J.~M. 1999, Science, 285, 862, \dodoi{10.1126/science.285.5429.862}

\bibitem[{{Nakariakov} \& {Verwichte}(2005)}]{2005LRSP....2....3N}
{Nakariakov}, V.~M., \& {Verwichte}, E. 2005, Living Reviews in Solar Physics,
  2, 3, \dodoi{10.12942/lrsp-2005-3}

\bibitem[{{Pascoe} {et~al.}(2017){Pascoe}, {Anfinogentov}, {Nistic{\`o}},
  {Goddard}, \& {Nakariakov}}]{2017A&A...600A..78P}
{Pascoe}, D.~J., {Anfinogentov}, S., {Nistic{\`o}}, G., {Goddard}, C.~R., \&
  {Nakariakov}, V.~M. 2017, \aap, 600, A78, \dodoi{10.1051/0004-6361/201629702}

\bibitem[{Pascoe {et~al.}(2022)Pascoe, Doorsselaere, \& Moortel}]{Pascoe_2022}
Pascoe, D.~J., Doorsselaere, T.~V., \& Moortel, I.~D. 2022, The Astrophysical
  Journal, 929, 101, \dodoi{10.3847/1538-4357/ac5e30}

\bibitem[{{Pascoe} {et~al.}(2020){Pascoe}, {Smyrli}, {Van Doorsselaere}, \&
  {Broomhall}}]{2020ApJ...905...70P}
{Pascoe}, D.~J., {Smyrli}, A., {Van Doorsselaere}, T., \& {Broomhall}, A.~M.
  2020, \apj, 905, 70, \dodoi{10.3847/1538-4357/abc69d}

\bibitem[{{Pesnell} {et~al.}(2012){Pesnell}, {Thompson}, \&
  {Chamberlin}}]{2012SoPh..275....3P}
{Pesnell}, W.~D., {Thompson}, B.~J., \& {Chamberlin}, P.~C. 2012, \solphys,
  275, 3, \dodoi{10.1007/s11207-011-9841-3}

\bibitem[{{R{\'e}gnier}(2007)}]{2007MmSAI..78..126R}
{R{\'e}gnier}, S. 2007, \memsai, 78, 126

\bibitem[{{R{\'e}gnier} \& {Priest}(2007)}]{2007A&A...468..701R}
{R{\'e}gnier}, S., \& {Priest}, E.~R. 2007, \aap, 468, 701,
  \dodoi{10.1051/0004-6361:20077318}

\bibitem[{{Roberts} {et~al.}(1983){Roberts}, {Edwin}, \&
  {Benz}}]{1983Natur.305..688R}
{Roberts}, B., {Edwin}, P.~M., \& {Benz}, A.~O. 1983, \nat, 305, 688,
  \dodoi{10.1038/305688a0}

\bibitem[{{Roberts} {et~al.}(1984){Roberts}, {Edwin}, \&
  {Benz}}]{1984ApJ...279..857R}
---. 1984, \apj, 279, 857, \dodoi{10.1086/161956}

\bibitem[{{Rosenberg}(1970)}]{1970A&A.....9..159R}
{Rosenberg}, H. 1970, \aap, 9, 159

\bibitem[{Sarkar {et~al.}(2016)Sarkar, Pant, Srivastava, \&
  Banerjee}]{sarkar2016transverse}
Sarkar, S., Pant, V., Srivastava, A., \& Banerjee, D. 2016, Solar Physics, 291,
  3269

\bibitem[{{Schatten} {et~al.}(1969){Schatten}, {Wilcox}, \&
  {Ness}}]{1969SoPh....6..442S}
{Schatten}, K.~H., {Wilcox}, J.~M., \& {Ness}, N.~F. 1969, \solphys, 6, 442,
  \dodoi{10.1007/BF00146478}

\bibitem[{Scherrer \& McKenzie(2017)}]{Scherrer_2017}
Scherrer, B., \& McKenzie, D. 2017, The Astrophysical Journal, 837, 24,
  \dodoi{10.3847/1538-4357/aa5d59}

\bibitem[{Su {et~al.}(2009)Su, van Ballegooijen, Lites, Deluca, Golub, Grigis,
  Huang, \& Ji}]{Su_2009}
Su, Y., van Ballegooijen, A., Lites, B.~W., {et~al.} 2009, The Astrophysical
  Journal, 691, 105, \dodoi{10.1088/0004-637x/691/1/105}

\bibitem[{Tian {et~al.}(2012)Tian, McIntosh, Wang, Ofman, Pontieu, Innes, \&
  Peter}]{Tian_2012}
Tian, H., McIntosh, S.~W., Wang, T., {et~al.} 2012, The Astrophysical Journal,
  759, 144, \dodoi{10.1088/0004-637x/759/2/144}

\bibitem[{{Tomczyk} \& {McIntosh}(2009)}]{2009ApJ...697.1384T}
{Tomczyk}, S., \& {McIntosh}, S.~W. 2009, \apj, 697, 1384,
  \dodoi{10.1088/0004-637X/697/2/1384}

\bibitem[{Tomczyk \& McIntosh(2009)}]{10.1088/0004-637x/697/2/1384}
Tomczyk, S., \& McIntosh, S.~W. 2009, The Astrophysical Journal, 697, 1384,
  \dodoi{10.1088/0004-637x/697/2/1384}

\bibitem[{Tomczyk {et~al.}(2007)Tomczyk, McIntosh, Keil, Judge, Schad, Seeley,
  \& Edmondson}]{doi:10.1126/science.1143304}
Tomczyk, S., McIntosh, S.~W., Keil, S.~L., {et~al.} 2007, Science, 317, 1192,
  \dodoi{10.1126/science.1143304}

\bibitem[{{Tomczyk} {et~al.}(2008){Tomczyk}, {Card}, {Darnell}, {Elmore},
  {Lull}, {Nelson}, {Streander}, {Burkepile}, {Casini}, \&
  {Judge}}]{2008SoPh..247..411T}
{Tomczyk}, S., {Card}, G.~L., {Darnell}, T., {et~al.} 2008, \solphys, 247, 411,
  \dodoi{10.1007/s11207-007-9103-6}

\bibitem[{{Uchida}(1970)}]{1970PASJ...22..341U}
{Uchida}, Y. 1970, \pasj, 22, 341

\bibitem[{{Van Doorsselaere} {et~al.}(2008){Van Doorsselaere}, {Nakariakov},
  {Young}, \& {Verwichte}}]{2008A&A...487L..17V}
{Van Doorsselaere}, T., {Nakariakov}, V.~M., {Young}, P.~R., \& {Verwichte}, E.
  2008, \aap, 487, L17, \dodoi{10.1051/0004-6361:200810186}

\bibitem[{{Wang} \& {Sheeley}(1992)}]{1992ApJ...392..310W}
{Wang}, Y.~M., \& {Sheeley}, N.~R., J. 1992, \apj, 392, 310,
  \dodoi{10.1086/171430}

\bibitem[{{Xue} {et~al.}(2016){Xue}, {Yan}, {Cheng}, {Yang}, {Su}, {Kliem},
  {Zhang}, {Liu}, {Bi}, {Xiang}, {Yang}, \& {Zhao}}]{2016NatCo...711837X}
{Xue}, Z., {Yan}, X., {Cheng}, X., {et~al.} 2016, Nature Communications, 7,
  11837, \dodoi{10.1038/ncomms11837}

\bibitem[{{Yan} \& {Sakurai}(2000)}]{2000SoPh..195...89Y}
{Yan}, Y., \& {Sakurai}, T. 2000, \solphys, 195, 89,
  \dodoi{10.1023/A:1005248128673}

\bibitem[{{Yang} {et~al.}(2020{\natexlab{a}}){Yang}, {Tian}, {Tomczyk},
  {Morton}, {Bai}, {Samanta}, \& {Chen}}]{2020ScChE..63.2357Y}
{Yang}, Z., {Tian}, H., {Tomczyk}, S., {et~al.} 2020{\natexlab{a}}, Science in
  China E: Technological Sciences, 63, 2357, \dodoi{10.1007/s11431-020-1706-9}

\bibitem[{{Yang} {et~al.}(2020{\natexlab{b}}){Yang}, {Bethge}, {Tian},
  {Tomczyk}, {Morton}, {Del Zanna}, {McIntosh}, {Karak}, {Gibson}, {Samanta},
  {He}, {Chen}, \& {Wang}}]{2020Sci...369..694Y}
{Yang}, Z., {Bethge}, C., {Tian}, H., {et~al.} 2020{\natexlab{b}}, Science,
  369, 694, \dodoi{10.1126/science.abb4462}

\end{thebibliography}
\bibliographystyle{aasjournal}
\end{document}